\documentclass[11pt]{article}
\usepackage{latexsym}
\usepackage{amsfonts}
\usepackage{amsmath, amsthm, amssymb}
\usepackage{graphicx}
\usepackage{fullpage}
\usepackage{setspace}
\usepackage{threeparttable}
\usepackage{longtable}
\usepackage{natbib}
\bibpunct{(}{)}{;}{a}{}{,}
\usepackage{dcolumn}
\newcolumntype{.}{D{.}{.}{-1}}
\newcolumntype{d}[1]{D{.}{.}{#1}}
\usepackage[top=1in, left=1in, right=1in,
bottom=1in]{geometry}
\usepackage{bm}
\usepackage{bbm}
\usepackage{booktabs}

\usepackage{comment}

\usepackage{color}
\usepackage[pdftex,
    colorlinks=true,
    linkcolor=blue,
    citecolor=blue,
    urlcolor=blue,
    filecolor=blue]{hyperref}

\usepackage{caption}
\usepackage{subcaption}

\usepackage{pgf,tikz}
\usetikzlibrary{positioning}
\usetikzlibrary{shapes}

\newcommand\indep{\protect\mathpalette{\protect\independenT}{\perp}}

\def\independenT#1#2{\mathrel{\rlap{$#1#2$}\mkern2mu{#1#2}}}

\newcommand{\E}{\ensuremath{\mathbb{E}}}
\newcommand{\R}{\ensuremath{\mathbb{R}}}

\theoremstyle{definition}
\newtheorem{assumption}{Assumption}

\begin{document}
\pagestyle{plain}

\newcommand{\blind}{0}

\newcommand{\tit}{A Weighting Framework for Clusters as Confounders in Observational Studies}

\if0\blind

{\title{\tit}
\author{Eli Ben-Michael\thanks{Carnegie Mellon University, 4800 Forbes Ave, Pittsburgh, PA, 15213, United States, Email: ebenmichael@cmu.edu}
\and Avi Feller\thanks{University of California, Berkeley, 2607 Hearst Avenue, Berkeley, CA, 94720, United States, Email: afeller@berkeley.edu}
\and Luke Keele\thanks{University of Pennsylvania, 3400 Spruce St, Philadelphia, PA, 19104, United States, Email: luke.keele@gmail.com}
}

\date{\today}

\maketitle
}\fi

\if1\blind
\title{\bf \tit}
\maketitle
\fi

\begin{abstract}
\noindent 

When units in observational studies are clustered in groups, such as students in schools or patients in hospitals, researchers often address confounding by adjusting for cluster-level covariates or cluster membership.
In this paper, we develop a unified weighting framework that clarifies how different estimation methods control two distinct sources of imbalance: global balance (differences between treated and control units across clusters) and local balance (differences within clusters).
We show that inverse propensity score weighting (IPW) with a random effects propensity score model --- the current standard in the literature --- targets only global balance and constant level shifts across clusters, but imposes no constraints on local balance.
We then present two approaches that target both forms of balance. 
First, hierarchical balancing weights directly control global and local balance through a constrained optimization problem.
Second, building on the recently proposed Generalized Mundlak approach, we develop a novel Mundlak balancing weights estimator that adjusts for cluster-level sufficient statistics rather than cluster indicators; this approach can accommodate small clusters where all units are treated or untreated.
Critically, these approaches rest on different assumptions: hierarchical balancing weights require only that treatment is ignorable given covariates and cluster membership, while Mundlak methods additionally require an exponential family structure.
We then compare these methods in a simulation study and in two applications in education and health services research that exhibit very different cluster structures. 
\end{abstract}

\begin{center}
\end{center}

\clearpage

\onehalfspacing

\section{Introduction}
In standard observational causal inference, researchers assume that there are no unmeasured confounders after adjusting for individual-level covariates alone. In many observational studies, however, individuals are clustered in groups, such as students nested in schools or patients nested in hospitals. In such cases, researchers can weaken this assumption by additionally adjusting for cluster-level covariates or for cluster membership itself. Consider an observational study evaluating a new surgical procedure, as we do in Section \ref{sec:app}. In addition to adjusting for patient-level covariates such as age, measures of health, and comorbidities, we might worry that hospital-level factors also influence both the treatment decision and the outcome. For example, the new procedure may be used more often in hospitals in which surgeons receive more advanced training. To capture this, we might want to additionally adjust for a hospital-level measure of surgical training. Alternatively, we might want to adjust for the admitting hospital itself, which could also capture other important aspects of ``unmeasured context'' \citep{Arpino:2011}. This simple example highlights some of the many trade-offs researchers face when adjusting for confounding in this setting. 

This paper reviews the many weighting estimators that have been developed to account for cluster-level confounding in observational causal inference.
Building on several reviews of model-based inverse propensity score weighting (IPW) in this clustered setting \citep{li2013propensity,chang2022propensity,fuentes2022causal}, we focus on adapting two new methodological developments to this area. 
First, we consider \textit{balancing weights}, a popular alternative to model-based IP weights that instead uses constrained optimization to directly find weights that balance covariates between groups \citep{benmichael_balancing_review,benmichael2021_lor}. 
We show that extant balancing weights and model-based IPW methods primarily target a combination of \textit{global balance}---covariate differences between treated and control units across clusters---and \textit{local balance}---covariate differences between treated and control units within clusters. Model-based IPW with a random cluster intercept, by far the most common approach in the literature, implicitly targets global balance alone and does not include any control of local balance. By contrast, both model-based IPW with random coefficients \citep{Arpino:2011} and hierarchical balancing weights \citep{benmichael2021_lor} target both global and local balance.  
A major challenge is that achieving both excellent global and local balance is often infeasible due to data sparsity; local balance within some clusters may be especially difficult to achieve due to limited cluster sample sizes. We show that several methods require dropping small clusters or clusters with nearly all treated or control units, which shifts the target estimand.

Second, we consider \textit{Generalized Mundlak Estimation} \citep{arkhangelsky2024fixed}, which adjusts for cluster-level sufficient statistics in place of adjusting for cluster membership itself. The key assumption is that treatment assignment is ignorable given individual-level information and cluster-level sufficient statistics (Assumption~\ref{a:exp_cluster_ignore}) . This is implied by ignorable treatment assignment given individual-level information and cluster membership, along with an additional assumption that the joint distribution of individual-level covariates and treatments follows an exponential family distribution. Unlike methods that adjust for cluster membership, the Mundlak approach can retain smaller clusters, albeit at the cost of imposing the stronger exponential family assumption that requires the cluster-level sufficient statistics to fully capture the relevant cluster-level information. We then propose \textit{Mundlak balancing weights}, a balancing weights estimator that also adjusts for cluster-level sufficient statistics. We outline two variants of this estimator, which constrain global balance in different ways, and argue that these are attractive estimators in practice.

Finally, we illustrate the trade-offs between these estimators with a simulation study and with two empirical illustrations with different cluster structures. The first application is an observational study of the impact of smaller class sizes on test scores, in which students are nested in schools with relatively small cluster sizes. The second application is an observational study of emergency general surgery for hepato-pancreato-biliary disorders, in which patients are nested within hospitals with many large clusters. We show that trade-offs differ in these two settings, and offer some suggestions for practice.


Our article is organized as follows. In Section~\ref{sec:prelim} we outline notation, estimands, and identification assumptions, and review common weighting estimation approaches for causal inference. In Section~\ref{sec:global_local}, we develop our unified framework for global and local balance. In Section~\ref{sec:estimation_clust_membership}, we use this framework to analyze extant methods that condition on cluster membership. In Section~\ref{sec:exp}, we use the framework to analyze methods that instead condition on cluster-level covariates. In Section~\ref{sec:inf}, we outline inference and bias correction. In Section~\ref{sec:sim} we conduct a simulation study. In Section~\ref{sec:app}, we analyze two different empirical applications.

\begin{table}[tbp]
\centering
\small
\begin{tabular}{@{}p{6cm}p{5cm}p{5cm}@{}}
\toprule
\textbf{Method} & \textbf{Balance Target} & \textbf{Main Limitation} \\
\midrule
\multicolumn{3}{l}{\textit{Methods that adjust for cluster membership (Assumption 2)}} \\
\midrule
Cluster-Specific Intercepts & Global only & No control over local balance \\
\addlinespace
Stratified Analysis & Local only & Must drop clusters with no treatment variation \\
\addlinespace
Random Coefficients & Global + Local  & Computationally difficult with convergence issues \\
\addlinespace
Hierarchical balancing weights & Global + Local  & Must drop clusters with all/no treated units \\
\addlinespace
\midrule
\multicolumn{3}{l}{\textit{Methods that adjust for cluster-level covariates  (Assumption 3; stronger than Assumption 2)}} \\
\midrule
Generalized Mundlak Estimator & Global + Local & Relies on model-based propensity score; Requires exponential family assumption and specification of sufficient statistics \\
\addlinespace
Mundlak Balancing Weights & Global + Local & Requires exponential family assumption and specification of sufficient statistics \\
\bottomrule
\end{tabular}
\caption{Overview of Weighting Methods for Cluster Adjustment}
\label{tab:methods_overview}

\end{table}

\section{Preliminaries}
\label{sec:prelim}
\subsection{Notation and problem setup}

We first outline our notation and problem setup, most of which is standard for describing an observational study with individualized treatment assignment.
We assume we have a sample of units $i=1,\dots,n$, drawn i.i.d. from some population, and denote  $Z_i$ as a binary treatment where $Z_{i} = 1$ corresponds to the treated condition, and $Z_{i} = 0$  corresponds to the control condition; $n_1$ denotes the number of units assigned to treatment, $n_0$ denotes the number assigned to control. We let $Y_i \in \R$ denote the observed outcome. We use $X_i \in \mathcal{X}$ to denote a vector of baseline covariates. To reduce notational complexity, we will drop the $i$ subscript when taking expectations. 

Next, we define some additional notation specific to the context where it is thought that group or cluster membership is related to treatment assignment; we use the terms \textit{group} or \textit{cluster} interchangeably to denote how units are nested. Let the group indicator $G_i \in \{1,\ldots,K\}$ denote membership for unit $i$ in that group. This notation might describe, for example, students in schools or patients in hospitals. We use $n_{1g}$ and $n_{0g}$ to denote the number of treated and control units in group $G_i = g$.

Denote the \emph{propensity score}, the probability of receiving treatment conditional on the baseline covariates, as $e(x) \equiv P(Z = 1 \mid X = x)$. And define the conditional expected outcome $m(z, x) \equiv \E[Y \mid Z = z, X = x]$.
We will also consider the propensity score and conditional expected potential outcome when additionally conditioning on group membership: $e(x, g) \equiv P(Z = 1 \mid X = x, G = g)$ and $m(z, x, g) \equiv \E[Y \mid Z = z, X = x, G = g]$.

\subsection{Assumptions and estimands}
\label{sec:assume}
We outline the key set of assumptions for an observational study, and how the presence of clusters changes the identification of treatment effects.
First, we make the standard Stable Unit Treatment Value Assumption \citep{Rubin:1986} that there are no  hidden forms of treatment and that  a subject's potential outcome is not affected by other subjects' treatment. We therefore rule out applications with clustered interference, where units interact within groups \citep[see, for example][]{hudgens2008toward}.
We can then define each unit's two potential responses, $(Y_{i}(1), Y_{i}(0)) \in \R^2$, corresponding to their response under treatment and control. 
These assumptions imply that the observed outcome is $Y_i = Z_iY_i(1) + (1 - Z_i) Y_i(0)$. 

Our focus will be on expected contrasts between potential outcomes or counterfactual states. In observational studies the target of inference is often the \emph{Average Treatment effect on the Treated} (ATT), $\tau \equiv \E[Y(1) - Y(0) \mid Z = 1] = \mu_1 - \mu_0$, where we denote $\mu_1 \equiv \E[Y(1) \mid Z = 1]$ and $\mu_0 \equiv \E[Y(0) \mid Z = 0]$.
Estimating $\mu_1$ is straightforward: the average observed outcome among the treated units is an unbiased estimate.
The difficulty then lies in estimating $\mu_0$, corresponding to the expected counterfactual outcome under control for the treated units.
We focus on the ATT for brevity, but our results generalize to other linear estimands, including the Average Treatment Effect (ATE) and the Average Treatment effect on the Overlap population \citep[ATO;][]{li2018balancing}.

\paragraph{Ignorability without cluster membership.}
In observational studies targeting the ATT without cluster-level information, the standard assumption is \emph{strong ignorability} \citep{Rosenbaum1983}. This combines (i) \emph{ignorable treatment assignment}---treatment is independent of the potential outcomes given the individual-level covariates---and (ii) \emph{overlap}---no individual deterministically receives treatment.
\begin{assumption}[Strong Ignorability]
  \label{a:ignore}
    $Y(1), Y(0) \indep Z | X$ and $e(x) < 1$ for all $x \in \mathcal{X}$
\end{assumption}
\noindent Note that other estimands have different overlap requirements, e.g. for the ATE we would require a stronger assumption that $0 < e(x) < 1$ for all $x \in \mathcal{X}$. Under Assumption \ref{a:ignore}, we can identify the average counterfactual outcome for the treated units as
\[
  \mu_0 = \E[m(0, X) \mid Z = 1] = \E\left[\frac{e(X)}{1 - e(X)} Y \mid Z = 0 \right].
\]

When cluster membership is thought to be relevant, analysts may also be interested in estimating the effect \emph{within cluster} $g$, e.g. the Conditional ATT (CATT),
\begin{equation}
    \label{eq:catt}
    \tau_g = \E[Y(1) - Y(0) \mid G = g, Z = 1] = \mu_{1g}-\mu_{0g},
\end{equation}
where similarly we denote $\mu_{1g} \equiv \E[Y(1) \mid G = g, Z = 1]$ and $\mu_{0g} \equiv \E[Y(0) \mid G = g, Z = 1]$. Again, estimating $\mu_{1g}$ is relatively straightforward: we can simply use the average outcome for treated units in cluster $g$. However, estimating $\mu_{0g}$ is again more difficult due to confounding.

\paragraph{Ignorability given cluster membership.}

In the presence of clusters, we can extend the standard ignorability assumption to include cluster membership. Specifically, we assume that treatment assignment satisfies strong ignorability given both the covariates $X$ and cluster membership $G$: 

\begin{assumption}[Strong Ignorability with Cluster Membership]
  \label{a:cluster_ignore}
    $Y(1), Y(0) \indep Z | X, G$ and $e(X, G) < 1$ for all $(x, g) \in \mathcal{X} \times \{1,\ldots,K\}$
\end{assumption}
Strong ignorability with cluster membership assumes that strong ignorability holds \emph{within clusters}, and not necessarily \emph{across clusters}.
In this setting we can view each cluster $g$ as a separate population with a separate observational study for which we estimate the CATT $\tau_g$ and then aggregate to obtain the overall ATT $\tau$.
Under this assumption we can identify $\mu_0$ and $\mu_{0g}$ as:
\begin{equation}
    \label{eq:identify_catt}
    \begin{aligned}
      \mu_0 & = \E[m(0, X, G) \mid Z = 1] = \E\left[\frac{e(X, G)}{1 - e(X,G)} Y \mid Z = 0\right],\\
    \mu_{0g} & = \E[m(0, X, g) \mid Z = 1, G = g] = \E\left[\frac{e(X, g)}{1 - e(X,g)} Y \mid Z = 0, G = g \right].  
    \end{aligned}
\end{equation}

By conditioning on cluster membership, the cluster ignorability assumption may be more palatable than the standard ignorability assumption, since it allows for the possibility that treatment assignment is confounded by unobserved cluster-level context. For example, consider comparing two different medical treatments. If we condition on the hospital where the patient receives treatment, we seek to only compare treated and control patients within the same hospital: the goal is to control for unobserved aspects of hospitals that contribute to the outcome,  holding all aspects of care associated with the hospital constant by comparing patients within the same hospital.

However, conditioning on cluster membership may make the overlap assumption more difficult to satisfy, since it requires overlap to hold for the treated and control comparison within \emph{each} cluster instead of across the entire study sample. In addition, in many empirical settings analysts will encounter clusters with small sample sizes, which means there are very few or no treated units, leading to (near) empirical overlap violations. In these settings, it will generally not be possible to nonparametrically estimate the CATT $\tau_g$ for such clusters. One solution is to remove small clusters from the data, especially if there is no treatment variation within those clusters. However, this changes the target population and the corresponding estimand to a particular local version of the ATT, which may not be well-defined.

\paragraph{Ignorability given cluster-level sufficient statistics.}

An alternative approach is to make a stronger assumption that allows us to pool information across clusters, while still allowing for cluster-level confounding. \citet{arkhangelsky2024fixed} show that Assumption~\ref{a:cluster_ignore} is equivalent to assuming that the potential outcomes are conditionally independent from treatment given the covariates and the \emph{empirical distribution} of the covariates and treatment in the cluster, i.e.,
    \[Y(1), Y(0) \indep Z \mid X, P_G,\]
where $P_g$ is the empirical distribution of $X$ and $Z$ in cluster $g$. This characterization is particularly useful if we place additional assumptions on the distribution of $X$ and $Z$. \citet{arkhangelsky2024fixed} show that if the distribution is an \emph{exponential family} distribution with sufficient statistic $S(x, z) \in \R^p$, then Assumption~\ref{a:cluster_ignore} implies that treatment assignment satisfies strong ignorability given the covariates and the average value of the sufficient statistic in the cluster, $\bar{S}_g = \frac{1}{n_g}\sum_{G_i = g} S(X_i, Z_i)$. We can formally write this assumption as:

\begin{assumption}[Exponential Family Cluster Ignorability]
  \label{a:exp_cluster_ignore}
    $Y(1), Y(0) \indep Z \mid X, \bar{S}_G$ and $e(X, \bar{S}_G) < 1$ for all $(x, \bar{s}) \in \mathcal{X}$.
\end{assumption}
\noindent Under this assumption, we can identify both $\mu_0$ and $\mu_{0g}$ in Equation \eqref{eq:identify_catt}, replacing the propensity score that conditions on cluster membership $e(X,G)$ with $e(X, \bar{S}_G)$, which instead conditions on the group-level sufficient statistics. See also \citet{he2018inverse}, who proposes an alternative identification approach that also conditions on cluster-level statistics. In this approach, the exponential family cluster ignorability Assumption~\ref{a:exp_cluster_ignore} allows the clusters to be ``anonymous'' in the sense that we do not need to know to which unit a cluster specifically belongs. Instead, if clusters have similar sufficient statistics, then we expect their relationship to the outcome to be similar as well.
As we will see, under this assumption it will be possible to estimate the ATT even with small clusters. Critically, however, Assumption~\ref{a:exp_cluster_ignore} is \emph{stronger} than Assumption~\ref{a:cluster_ignore}: while Assumption~\ref{a:cluster_ignore} is nonparametric in cluster membership, Assumption~\ref{a:exp_cluster_ignore} additionally requires that the joint distribution of $(X, Z)$ within clusters belongs to an exponential family and that the chosen sufficient statistics $\bar{S}_g$ fully capture all relevant cluster-level information. If the exponential family assumption is violated or the sufficient statistics are misspecified, the resulting estimates may be biased even when Assumption~\ref{a:cluster_ignore} holds.

\subsection{Estimating inverse propensity score weights}
Given the assumptions above and the identification formula in Equation \eqref{eq:identify_catt}, the next step is to estimate the corresponding ATT weights, $\gamma_i = e(x_i, \cdot)/(1-e(x_i, \cdot))$, where the second argument is either $g_i$ or $\bar{s}_{g_i}$. We focus on two common approaches to estimating these weights: the traditional modeling approach to inverse propensity score weighting and the recently popular balancing weights approach. Here we briefly review these methods with individual-level only data; in the next sections, we explore extending these methods to also adjust for cluster-level data.

\paragraph{Model-based IP weights.} The traditional, model-based approach to IPW \citep{Rosenbaum1987} first estimates the propensity score $\hat{e}(x_i)$ and then plugs in the estimated $\hat{e}$ into the known identification formula in Equation \eqref{eq:identify_catt}.
By far the most common approach is to fit a logistic regression model of treatment assignment $Z$ on covariates. With individual-only covariates, this is $e(x_i) = \text{logit}^{-1}(\theta \cdot x_i)$, with corresponding estimated propensity scores $\hat{e}(x_i) = \text{logit}^{-1}(\hat\theta \cdot x_i)$. The model-based IP weights are then:
$
\hat{\gamma}_i^{ipw} = \hat{e}(x_i) / (1 - \hat{e}(x_i )).
$
A key issue in practice is that the weights $\hat{\gamma}_i^{ipw}$ only balance the covariates $X$ indirectly and can often lead to poor finite-sample balance: i.e., $\sum(1-Z_i)\hat{\gamma}_i^{ipw}X_i \neq \sum Z_i X_i$ \citep{Chattopadhyay2020,ben2023using}. As we will see below, this problem is often exacerbated when seeking to balance cluster-level data as well.

\paragraph{Balancing weights.} An alternative approach, known as \textit{balancing weights} in causal inference and \textit{calibration} in the surveys literature, instead seeks to directly find weights that balance covariates \citep{Hainmueller2011,zubizarreta2015stable,Zhao2016a, Zhao2019,Wang2020, Chattopadhyay2020}. The resulting weights often have better finite sample behavior than traditional model-based IP weights; see \citet{benmichael_balancing_review} for a review.

These approaches begin by evaluating the bias of a generic weighting estimator:
\begin{equation}
  \label{eq:bias_nocluster}
  \text{Bias} = \frac{1}{n_1}\sum_{i=1}^n Z_i m(0, X_i) - \frac{1}{n_1}\sum_{i=1}^n (1 - Z_i) \gamma_i m(0, X_i).
\end{equation}
This bias is the imbalance in a particular function of the covariates: the expected potential outcome under control given the covariates, $m(0, x)$.
Balancing weights approaches then consider a (parametric or non-parametric) model for $m(0,x)$ and create an imbalance metric by computing the worst-case bias across the range of potential models.
Finally, these optimization-based approaches find weights that minimize a combination of the imbalance metric --- a stand-in for the bias --- and a measure of the dispersion of the weights --- a stand-in for the variance. There are many possible choices of imbalance metric and dispersion parameter. Following \citet{ben2024estimating}, much of our discussion will focus on the choice of imbalance metric in particular.

Importantly, these balancing weights approaches
\emph{indirectly} estimate a propensity score model. We can see this using the identification result that $\gamma = e(x)/(1-e(x))$ --- and therefore $e(x) = \gamma/(1+\gamma)$.
More formally, one can show that optimization-based balancing weights approaches are dual to calibrated propensity score estimation approaches \citep{Zhao2019}. Thus, we can view balancing weights approaches as estimating a propensity score model with a slightly different loss function than the usual maximum likelihood estimators. Moreover, balancing weights often outperform IP weights estimated from logistic regression, sometimes have double-robustness guarantees when using the weights alone, and can be de-biased with an outcome model to give double-robustness properties and enable asymptotically valid inference in high dimensional or non-parametric settings \citep{Chattopadhyay2020, Wang2020, zhou2020propensity, Hirshberg2019_amle, ben2023using, benmichael2022}.

\section{A global and local balance framework}
\label{sec:global_local}

We next decompose the bias of weighting estimators for the ATT in the presence of clusters.
As we have discussed, the key challenge is estimating the mean counterfactual outcome for the treated units, $\mu_0 = \mathbb{E}[Y(0) \mid Z = 1]$.
For weights $\gamma_i$, we estimate $\mu_0$ with
\[
\hat{\mu}_0 = \frac{1}{n_1}\sum_{i=1}^n \gamma_i Y_i (1 - Z_i).
\]
\noindent We can decompose the bias in estimating $\mu_0$ into two components: a \emph{global balance} component and a \emph{local balance} component. To do so, we first note that we can write any outcome model under control $m(0, x, g)$ as
\begin{align*}
m(0, x, g) &= \eta_g + \underbrace{\E[Y - \eta_g \mid Z = 0, X=  x]}_{f(x)} + \underbrace{\E[Y \mid Z = 0, X = x, G = g] - \E[Y \mid  Z = 0, X = x]}_{\delta(x, g)},
\end{align*}
where $\eta_g$ is a cluster-specific fixed effect, $f(x)$ is a common component of the model shared across clusters, and $\delta(x, g)$ is an interaction component specific to cluster $g$. With this notation, we can decompose the bias in $\hat{\mu}_0$ (conditional on $X$, $G$, and $Z$) into three terms:
\begin{equation}
  \label{eq:error_global}
  \begin{aligned}
      \text{Bias} & = \sum_g \eta_g \underbrace{\left(\frac{1}{n_1}\sum_{G_i = g}(1 - Z_i) \gamma_i  - \frac{n_{1g}}{n_1}\right)}_\text{average-to-one-within-cluster} + \underbrace{\frac{1}{n_1} \sum_{i=1}^n (1 - Z_i) \gamma_i f(X_i) - \frac{1}{n_1} \sum_{i=1}^n Z_i f(X_i)}_\text{global balance}  \\
      & + \sum_{g} \underbrace{\frac{1}{n_1}\sum_{G_i = g} (1  -Z_i) \gamma_i \delta(X_i, g) - \frac{1}{n_1}\sum_{G_i = g} Z_i \delta(X_i, g) }_\text{local balance}.
  \end{aligned}
\end{equation}
The first term arises due to the presence of cluster-specific fixed effects. This can be viewed as a cluster-specific intercept shift in the bias. The presence of this term can lead to bias if the sum of the weights for control units in each cluster does not equal the number of treated units in that cluster; this term is therefore closely linked to normalization. The second term captures the \emph{global balance} component, the difference  in the common component of the outcome model $f(x)$ between treated and control units after weighting. Poor global balance implies that there are systematic difference between treated and control units \emph{across clusters}. That is, bias arises due to systematic differences between treated and control units without respect to clusters. Finally, the third term captures \emph{local balance}: the difference in the interaction component $\delta(x, g)$ between treated and control units within each cluster $g$. If local balance is poor for a specific cluster, then there are systematic differences between treated and control units \emph{within that cluster}.

\subsection{Restricting functions to balance}
\label{sec:restrict}
Equation~\eqref{eq:error_global} is non-parametric in the sense that it does not require a specific functional form for the outcome model $m(0, x, g)$. That said, it is often helpful to consider a specific functional form for the outcome model, or, equivalently to specify functions to balance. To that end, we introduce one additional piece of notation.
We use $\phi(x)$ to denote features of the covariates $x$ --- i.e. transformations of $x$ --- used to account for non-linearity. Possible transformations include polynomials in $x$ or interactions between covariates. Selection of such features can be done based on \emph{a priori} information or through data-driven methods \citep{ben2024estimating, jin_cross-balancing_2025, shen_forest_2025}.

To make progress, we focus on the setting in which the outcome model is assumed to be linear in features $\phi(x)$. This remains a very general formulation, nesting a wide class of flexible outcome models, including many trees, forests, and neural networks, as well as kernels and RKHS regression, with a possibly infinite basis for $\phi(x)$ \citep{benmichael_balancing_review}. Under linearity, we can then write the model with features  $\phi(x) \in \R^d$ as:
$$m(0, x, g) = \eta_g + (\beta + \delta_g) \cdot \phi(x),$$
where $\eta_g$ is a cluster-specific fixed effect, $\beta$ are common coefficients across clusters and $\delta_g$ is a cluster-specific interaction term. 

Now, we can rewrite the bias decomposition in Equation~\eqref{eq:error_global} directly in terms of (coefficient-weighted) covariate balance: 
\begin{equation}
  \label{eq:error_global_linear}
  \begin{aligned}
      \text{Bias} & =~~ \sum_g \eta_g \underbrace{\left(\frac{1}{n_1}\sum_{G_i = g}(1 - Z_i) \gamma_i  - \frac{n_{1g}}{n_1}\right)}_\text{average-to-one-within-cluster} ~~+~~ \beta \cdot \underbrace{\left(\frac{1}{n_1} \sum_{Z_i=0}^n \gamma_i \phi(X_i) - \frac{1}{n_1}\sum_{Z_i = 1 }^n\phi(X_i)\right)}_\text{global balance}  \\
      & +~~ \sum_{g} n_{1g} \delta_g \cdot \underbrace{\left(\frac{1}{n_{1g}}\sum_{G_i = g, Z_i = 0} \gamma_i \phi(X_i) - \frac{1}{n_{1g}}\sum_{G_i=g, Z_i = 1}\phi(X_i)\right)}_\text{local balance}.
  \end{aligned}
\end{equation}
Now we can see that the global balance term is the difference in the average features $\phi(x)$ \emph{across clusters} between treated and control units after weighting, and the local balance term is the difference in the average features $\phi(x)$ \emph{within each cluster} between treated and control units after weighting.
Therefore, when constructing weighting estimators with the intent of controlling for cluster membership, we want to achieve \emph{both} good global balance between treated and control units across clusters and good local balance between treated and control units within each cluster. 

The relative importance of local and global balance for estimating the overall ATT is controlled by the level of similarity in the outcome process across groups. In the extreme case where the outcome process does not vary across clusters --- i.e., $\delta_g = 0$ for all $g$ in the linear model --- then controlling global balance and ensuring that the weights average to one within each cluster is sufficient to control the bias.
If furthermore the cluster-level fixed effects are all the same -- i.e. $\eta_g = \eta$ for all $g$ --- then cluster membership does not contribute to the outcome at all, corresponding to a setting where the strong ignorability assumption without clusters (Assumption~\ref{a:ignore}) holds. Here, there would be no need to control for schools or hospitals. In the other extreme, where the outcome model varies substantially across clusters --- e.g., $\|\delta_g\|$ is large for all $g$ in the linear model --- there is some important component of the outcome that is related to schools, hospitals, or other types of clusters. In such cases we should primarily seek to control local imbalance within each cluster in order to reduce bias.

This bias decomposition motivates a general goal for observational studies with clusters: we seek to eliminate imbalance between treated and control units at both the global and local level. However, even if it is possible to exactly balance units at the global level, balancing at the local level may be difficult in practice due to data sparsity.

\section{Methods that adjust for cluster membership}
\label{sec:estimation_clust_membership}

We first consider adjustments that condition on cluster membership only, rather than adjust for cluster-level covariates. Under this approach, investigators can target global balance only, local balance only, and both global and local balance together. Here, we review each strategy and outline how standard practice fits under each rubric to clarify the balance assumptions that are being used.

\subsection{Model-based IP weights for global and local balance}
\label{sec:model-ipw-global-local}

We begin by reviewing implementations of model-based IP weights that (indirectly) target a combination of global and local balance. We focus on three main approaches here: including a cluster-specific intercept, conducting an analysis stratified by cluster membership, and including random coefficients. We turn to the balancing weights analogue next.

There are many additional IPW approaches that do not fit neatly into this setup.  
A number of papers have explored using more flexible methods for propensity scores while incorporating cluster effects, including via GRF methods, BART, GBMs, and latent class modeling \citep{kim2015multilevel,suk2021random,chang2022flexible,salditt2023parametric}.
In a different direction, \citet{lee2021partially} proposes to partially pool similar small clusters together into larger groups.

\paragraph{Targeting only global balance: cluster-specific intercepts.}

The main weighting approach to targeting global balance while allowing for level differences across clusters is to include a cluster-specific intercept in the propensity score model: $e(x, g) = f(\alpha_g + \theta \cdot \phi(x))$ for some link function $f$; see \citet{Arpino:2011,thoemmes2011use}. Returning to the bias decomposition in Equations~\eqref{eq:error_global} and \eqref{eq:error_global_linear}, the global approach assumes that there is no interaction component, i.e. $\delta(x, g) = 0$ for all $x$ and $g$; in the linear model, this corresponds to assuming that $m(0, x, g) = \eta_g + \beta \cdot \phi(x)$. Thus, this global approach does not attempt to balance unit-level covariates within each cluster, and any imbalance of this form is left unaccounted for. As such, this method targets only two aspects of the bias (i) the average-to-one term within each cluster (from the cluster-specific fixed effect);\footnote{Unstandardized logistic regression propensity score IPW only weakly guarantees balance as the sample size grows to infinity. Standardizing weights within each cluster, however, ensures the average-to-one-within-cluster constraint and zeros out the contribution of the fixed effects to the bias.
}
and (ii) the global balance term (from the common coefficients).

There are two ways to implement cluster-specific intercepts in the propensity score model: fixed and random effects.
The \textit{fixed effects} approach simply adds a dummy variable or fixed for each cluster in the logistic regression. Unfortunately, estimating a logistic propensity score model with cluster-specific intercepts (e.g., by including $G-1$ indicators) tends to run afoul of the incidental parameter problem \citep{neyman1948consistent,lancaster2000incidental}, and can lead to poor finite sample behavior.
The \textit{random effects} or \textit{random intercepts} approach instead partially pools the group-specific intercepts in a hierarchical model \citep{Arpino:2011,thoemmes2011use}; cluster intercepts for clusters with large variances will be shrunk more toward the sample average than those with small variance.\footnote{For alternative implementations, see, e.g., \citet{yang2018propensity}, who implement a covariate balancing propensity score approach in this vein.} 
Importantly, this approach regularizes the cluster fixed effects in Equation \eqref{eq:error_global} and so approximately satisfies the average-to-one constraint within each cluster.  We term this approach random-intercept IPW or (RI-IPW).  This approach is most widely used in applied work and has been frequently studied and reviewed in the literature \citep{li2013propensity,salditt2023parametric,chang2022propensity}

Notably, including a cluster-specific intercept does not target local balance because the propensity score model does not include an interaction between cluster membership and individual-level covariates. As a result, there is no control over local imbalance within clusters, even in very large samples. Thus, if there are strong imbalances within clusters, using a global approach alone will not control local balance and so will not be sufficient to control the bias. 

\paragraph{Targeting only local balance: stratified analysis.}

At the other extreme, we can estimate a propensity score model with all cluster-specific interactions, i.e. 
$e(x, g) = f(\alpha_g + \zeta_g \cdot \phi(x))$.
That is, the propensity score model would need to include the cluster fixed effect \emph{and} an interaction between each feature and the cluster indicator.
This is equivalent to estimating the propensity score separately within each cluster via a completely stratified analysis; it is 
typically infeasible in practice with a logistic regression propensity score model, which will typically exhibit ``complete separation'' \citep{gelman2008separation}. 

Even if the fully-interacted model is feasible to estimate, there are two main drawbacks to this approach.
First, local balance may be poor in clusters with small sample sizes: some clusters may have a very small pool of control units or it may simply be difficult to balance covariates well within some groups.
Second, this analysis entirely ignores global balance: if local balance is poor within each cluster, this will lead to poor global balance, even if it is in principle possible to achieve good global balance.
Alternatively, in many applications, there may be clusters with few or no treated units. One can discard these clusters, but this changes the estimand and may not correspond to a well-defined population of interest.

\paragraph{Targeting both local and global balance: random coefficients.}

Finally, we can target both local and global balance simultaneously by partially pooling across clusters via random intercepts and coefficients in the propensity score model:  $e(x, g) = f(\alpha_g + (\theta + \zeta_g) \cdot \phi(x))$, where $\alpha_g$ and $\zeta_g$ have a hierarchical component \citep{Arpino:2011,thoemmes2011use}. That is, the propensity score estimate for each unit will now have a covariate-specific component that varies with clusters, albeit one regularized across clusters.

Again, there are drawbacks to this approach. Typically, the goal is to locally balance \emph{all} unit-level covariates, which may require adding many random coefficient terms to the model. Estimating a large number of random coefficient terms is often computationally infeasible or requires special algorithms \citep{gao2020estimation}. The alternative is to only include a few random coefficient terms, but limiting the number of random coefficients targets local balance for only a limited set of covariates. In practice, it can be difficult to estimate a propensity score model with even a single random coefficient in larger samples. Moreover, for many data sets, convergence for these models is difficult if not impossible. 

\subsection{Balancing weights for global and local balance} 
\label{sec:bal_wts_global_local}

\citet{ben2024estimating} develop an alternative \textit{balancing weights} approach that estimates weights that target both global and local balance but allows for trade-offs between these goals. In this framework, the goal is to find weights that directly control the three terms in the bias decomposition in Equation~\eqref{eq:error_global_linear}. Importantly, this approach nests multilevel calibration estimators, such as \citet{yang2018propensity} and \citet{kim2017causal_nonignorable} \citep[see][for review]{fuentes2022causal}, which impose exact balance constraints in place of the approximate constraints we consider here.

The first step is to posit a model class for the shared common component $f(x)$ and the interaction components $\delta(x, g)$, i.e. $f \in \mathcal{F}$ and $\delta(\cdot, g) \in \Delta$ for all $g$. We can then find weights that minimize the worst case bias across these two model classes. This measures the global and local balance terms as
\begin{align*}
  \text{global balance} & = \sup_{f \in \mathcal{F}} \left|\frac{1}{n_1} \sum_{Z_i=0}^n \gamma_i f(X_i)- \frac{1}{n_1}\sum_{Z_i = 1 }^nf(X_i)\right|,\\
  \text{local balance}(g) & = \sup_{\delta(\cdot, g) \in \Delta} \left|\frac{1}{n_{1g}}\sum_{G_i = g, Z_i = 0} \gamma_i \delta(X_i, g) - \frac{1}{n_{1g}}\sum_{G_i=g, Z_i = 1}\delta(X_i, g)\right|.
\end{align*}

To operationalize this, we can use the linear-in-$\phi(x)$ model discussed in Section~\ref{sec:restrict}.
To do so, we include an additional restriction that the coefficients are bounded: the model class for the shared common component is $\mathcal{F} =  \{\beta\cdot\phi(x) \mid \|\beta\|_2 \leq B\}$ and the model class for the cluster-specific interaction components is $\Delta = \{\delta_g \cdot \phi(x) \mid \|\delta_g\|_2 \leq D\}$. The resulting imbalance terms are:
\begin{align*}
  \text{global balance} & = B\left\|\frac{1}{n_1} \sum_{Z_i=0}^n \gamma_i \phi(X_i)- \frac{1}{n_1}\sum_{Z_i = 1 }^n\phi(X_i)\right\|_2,\\
  \text{local balance}(g) & = D\left\|\frac{1}{n_{1g}}\sum_{G_i = g, Z_i = 0} \gamma_i \phi(X_i) - \frac{1}{n_{1g}}\sum_{G_i=g, Z_i = 1}\phi(X_i)\right\|_2,
\end{align*}
where $B$ is the maximum magnitude of the coefficients for the global component and $D$ is the magnitude of the coefficients for the interaction component.\footnote{Here we have used the Euclidean norm but other norms are also possible and readily implementable.}

Often we might expect that the common component of the outcome model is much larger than the interaction component, i.e. $B \gg D$.  That is, it may be the case that many of the local balance terms are small or close to zero. In such cases, it is natural to prioritize the global balance term over the local balance term. For example, \citet{benmichael2021_lor} and \citet{ben2024estimating} find weights that optimize for local balance while ensuring exact global balance across the sample. This leads to the following quadratic programming objective function:
\begin{equation}
    \label{eq:primal}
    \begin{aligned}
        \min_{\gamma} \quad & \sum_{g=1}^K \left\|\frac{1}{n_{1g}}\sum_{G_i = g, Z_i = 0} \gamma_i \phi(X_i) - \frac{1}{n_{1g}}\sum_{G_i = g,Z_i = 1} \phi(X_i)\right\|_2^2 \;\; + \;\; \frac{\lambda}{n_{1g}^2}\sum_{G_i=G,Z_i=0} \gamma_i^2\\[1.2em]
        \text{subject to }\quad  & \frac{1}{n_1}\sum_{Z_i = 0} \gamma_i \phi(X_i) = \frac{1}{n_1}\sum_{Z_i = 1}\phi(X_i)\\[1.2em]
        & \frac{1}{n_{1g}}\sum_{G_i = G, Z_i = 0} \gamma_i = 1, \qquad\qquad\qquad \gamma_i  \geq 0 \;\;\;\;\; \forall i=1,\ldots,n.
    \end{aligned}
\end{equation}

This objective function minimizes the local imbalance within each cluster $g$ while constraining the weights to have \emph{exact global balance} over the entire sample, as well as satisfying an average-to-one constraint within each cluster. There is also a regularization term that penalizes larger weights---and so penalizes high variance weights. The amount of regularization is controlled by the hyperparameter $\lambda$:  larger values of $\lambda$ allows for increased weight complexity and enforces a less strict balance constraint; smaller values of $\lambda$ do the opposite.

A major challenge with this approach is that it can be difficult to achieve good local balance within each cluster, especially when the clusters are small. The issue is essentially one of dimensionality: when including interactions, the number of balance constraints increases from $d$, the number of covariates, to $d \times K$, the number of covariates times the number of clusters. That is, while this approach indeed targets local balance in a principled way, achieving good balance on everything may be difficult with so many terms to balance. Finally, as with other methods we consider above, practical overlap violations are more likely in small clusters: there may be some small clusters where almost every unit within the cluster is treated or is not treated.
Again, the primary remedy for this is to drop small clusters from the analysis, which is often undesirable.


\section{Methods that adjust for cluster-level covariates}
\label{sec:exp}

An alternative approach is to use Assumption~\ref{a:exp_cluster_ignore} to reduce the dimensionality of the local balance constraints by using cluster-level sufficient statistics $\bar{S}_G$ in place of fixed effects for cluster membership. For example, \citet{arkhangelsky2024fixed} propose balancing on cluster-level averages of the unit-level covariates, the proportion treated in each cluster, and the interaction between these two measures instead of using the cluster fixed effects. They refer to this approach as \textit{Generalized Mundlak Estimation}, in reference to the fixed-effect estimator in \citet{mundlak1978pooling}. As we will see, when the dimensionality of the sufficient statistics is small relative to the number of clusters, this will reduce the number of balance constraints and make achieving good local balance more feasible. We argue that this approach is essentially an alternative form of regularization over the large number of cluster-specific interaction terms.

It is important to emphasize that the practical benefits of the Mundlak approach come at the cost of invoking Assumption~\ref{a:exp_cluster_ignore}, which is strictly stronger than Assumption~\ref{a:cluster_ignore}.
Methods in Section~\ref{sec:estimation_clust_membership} that adjust for cluster membership are valid under the weaker Assumption~\ref{a:cluster_ignore} and so do not make assumptions on the distribution of the covariates within the clusters, nor the form of cluster-level context. However, they may suffer from poor local balance due to small clusters.
The generalized Mundlak approach can achieve better local balance even with small clusters, but requires the additional exponential family structure to hold. Researchers should carefully consider whether this additional assumption is plausible in their application.

It is particularly important to accurately characterize the functional form of the sufficient statistics $S(x, z)$. Higher dimensional sufficient statistics that include interactions, non-linear transformations, or higher-order terms will result in a weaker Assumption~\ref{a:exp_cluster_ignore}, as the resulting exponential family distribution will be more flexible and can capture a wider range of cases. However, as we will see, higher dimensional sufficient statistics will also lead to more local balance constraints, which may make achieving good local balance more difficult in practice. In this way, the choice of sufficient statistics $S(x, z)$ is similar to the choice of the features $\phi(x)$ discussed in Section~\ref{sec:restrict}, and similar data-driven approaches may be useful to select these sufficient statistics.

\subsection{A Mundlak framework}
\label{sec:mundlak_setup}
Under Assumption~\ref{a:exp_cluster_ignore}, we consider the expected outcome conditional on treatment, the covariates, and the cluster-level sufficient statistics, $m(z, x, s) = \E[Y \mid Z = z, X = x, \bar{S}_G = s]$, and we can write a similar decomposition as in Section~\ref{sec:global_local} with the following equation:
\[
m(0, x, s) = \eta(s) + f(x) + \tilde{\delta}(x, s),
\]
where $\eta(s)$ is a function of the cluster-level sufficient statistics, $f(x)$ is a common component of the model shared across clusters, and $\tilde{\delta}(x, s)$ is an interaction component that depends on both the cluster-level sufficient statistics and the covariates. Here, the function $\eta(s)$ plays the role of the cluster-specific fixed effects in the previous section, but now it is a function of the cluster-level sufficient statistics rather than being unique for each cluster. Similarly, the interaction term $\tilde{\delta}(x, s)$ is now a function of the cluster-level sufficient statistics rather than cluster membership. Next, we can decompose the bias in estimating $\mu_0$ similarly to Equation~\eqref{eq:error_global} as 
\begin{equation}
  \label{eq:error_global_suff}
  \begin{aligned}
      \text{Bias} & = \underbrace{\frac{1}{n_1}\sum_{i=1}^n (1 - Z_i) \eta(\bar{S}_{G_i}) \gamma_i  - \frac{1}{n_1}\sum_{i=1}^n Z_i \eta(\bar{S}_{G_i})}_\text{global cluster balance} + \underbrace{\frac{1}{n_1} \sum_{i=1}^n (1 - Z_i) \gamma_i f(X_i) - \frac{1}{n_1} \sum_{i=1}^n Z_i f(X_i)}_\text{global individual balance}  \\
      & + \underbrace{\frac{1}{n_1} \sum_{i=1}^n (1 - Z_i) \gamma_i \tilde{\delta}(X_i, \bar{S}_{G_i}) - \frac{1}{n_1} \sum_{i=1}^n Z_i \tilde{\delta}(X_i, \bar{S}_{G_i})}_\text{local balance}.
  \end{aligned}
\end{equation}
\noindent Comparing this decomposition to one in  Equation~\eqref{eq:error_global}, we see that cluster membership $G_i$ is no longer directly relevant for the bias. Rather, the cluster-level sufficient statistics $\bar{S}_{G_i}$ alone capture the cluster-level information. In this approach, under Assumption~\ref{a:exp_cluster_ignore}, the clusters are treated as ``anonymous'' in the sense that we do not need to know which clusters contain which specific units. Instead, if clusters are similar in terms of their sufficient statistics as measured across all units in the cluster, then we expect their relationship to the outcome to be similar as well. We also see that it is enough to ensure that the weights balance the cluster-level sufficient statistics $\bar{S}_{G_i}$ between treated and control units across clusters, rather than requiring that the weights average to one within each cluster:
under Assumption~\ref{a:exp_cluster_ignore}, there is nothing unique about cluster membership that is not captured by the cluster-level sufficient statistics. Below, we consider an alternative constraint that the weights average to one within each cluster.

Thus, the primary benefit of conditioning on cluster-level sufficient statistics instead of on cluster membership is that we can achieve better local balance. The interaction term $\tilde{\delta}(x, s)$ is a function of the cluster-level sufficient statistics $\bar{S}_G$ rather than cluster membership $G$. We can therefore treat the interaction between the individual level covariates $X$ and the cluster-level sufficient statistics $\bar{S}_G$ as a new covariate---and can target local balance in terms of this new interaction covariate, rather than by trying to target local balance for all covariates within each cluster. However, this simplification is only valid under the stronger Assumption~\ref{a:exp_cluster_ignore}; if the sufficient statistics do not fully capture cluster-level confounding, bias may remain even with excellent balance on these terms.

\subsection{Mundlak balancing weights}
\label{sec:mundlak_bal_wts}
We now outline a balancing weights estimator to operationalize the Mundlak setup in Section 
\ref{sec:mundlak_setup}. This differs from \citet{arkhangelsky2024fixed}, who propose an estimation strategy using model-based propensity scores as well as an augmented approach that incorporates an outcome model for additional bias correction.\footnote{As in our discussion above, we could also fit a propensity score model with a random intercept or random coefficient terms, though \citet{arkhangelsky2024fixed} do not explicitly outline such a procedure.}

First, let $\psi(x, s)$ capture the relevant interaction terms between the individual-level covariates $x$ and the cluster-level sufficient statistics $s$. We can then extend the hierarchical balancing weights setup in Section \ref{sec:bal_wts_global_local} to balance $\psi(x,s)$:
\begin{equation}
    \label{eq:primal_suff}
    \begin{aligned}
        \min_{\gamma} \quad &  \left\|\frac{1}{n_1} \sum_{i=1}^n (1 - Z_i) \gamma_i \psi(X_i, \bar{S}_{G_i}) - \frac{1}{n_1} \sum_{i=1}^n Z_i \psi(X_i, \bar{S}_{G_i})\right\|_2^2 + \frac{\lambda}{n_{1}^2}\sum_{Z_i=0} \gamma_i^2\\[1.2em]
        \text{subject to }\quad  & \frac{1}{n_1}\sum_{Z_i = 0} \gamma_i \bar{S}_{G_i} = \frac{1}{n_1}\sum_{Z_i = 1}\bar{S}_{G_i}\\[1.2em]
        & \frac{1}{n_1}\sum_{Z_i = 0} \gamma_i \phi(X_i) = \frac{1}{n_1}\sum_{Z_i = 1}\phi(X_i)\\[1.2em]
        & \gamma_i  \geq 0 \;\;\;\;\; \forall i=1,\ldots,n.
    \end{aligned}
\end{equation}

Comparing to the optimization problem in Equation~\eqref{eq:primal}, we see that there is still an exact global balance constraint for the individual-level covariates, as well as a variance penalty on the weights. As before, $\lambda$ serves as a hyperparamter that controls the trade-off between bias reduction and weight stability.
However, the local balance term is now a function of the interaction between the individual-level covariates $X_i$ and the cluster-level sufficient statistics $\bar{S}_{G_i}$. This change to the local balance measure is particularly useful when there are near empirical overlap violations within clusters, which may occur when there are small clusters. Moreover, this form of local balance does not require balance within each cluster, and so it is feasible in settings where some clusters only have control or treated units. As such, there is no need to remove smaller clusters or clusters with few treated or control units. 

More broadly, this local constraint can be substantially easier to balance when there are many clusters and few units within each cluster. If we do not invoke Assumption~\ref{a:exp_cluster_ignore}, local balance implies that there are $d \times K$ balance constraints, where $d$ is the dimension of $\phi(x)$ and $K$ is the number of clusters. In contrast, by invoking Assumption~\ref{a:exp_cluster_ignore}, we can reduce the number of balance constraints to the dimensionality of $\psi(x, s)$, which will typically be smaller than $d \times K$. This reduction in constraints is the key practical advantage of the Mundlak approach. However, this advantage comes at a cost: the approach requires the stronger Assumption~\ref{a:exp_cluster_ignore} rather than merely Assumption~\ref{a:cluster_ignore}. Specifically, researchers must assume that (i) the joint distribution of $(X, Z)$ within clusters follows an exponential family, and (ii) the chosen sufficient statistics $\bar{S}_g$ capture all cluster-level confounding. If either condition fails, the Mundlak approach may yield biased estimates even when methods that directly condition on cluster membership would remain valid.


\paragraph{Average-to-one constraint vs.\ global balance.} As we discuss in Section \ref{sec:mundlak_setup}, the global balance constraints on $\bar{S}_{G_i}$ in Equation \eqref{eq:primal_suff} control the bias of the global cluster balance term in Equation \eqref{eq:error_global_suff}.
However, it is also sufficient to constrain weights to average to one within each cluster. To see this, note that we can write the global balance cluster term to be in the same form as the average-to-one term in Equation~\eqref{eq:error_global}:
\[
\frac{1}{n_1}\sum_{i=1}^n (1 - Z_i) \eta(\bar{S}_{G_i}) \gamma_i  - \frac{1}{n_1}\sum_{i=1}^n Z_i \eta(\bar{S}_{G_i}) = \frac{1}{n_1}\sum_g n_{1g} \eta(\bar{S}_g)\left(\frac{1}{n_{1g}}\sum_{G_i = g}(1 - Z_i) \gamma_i  -1\right).
\]
Thus, we can consider a variation of the optimization problem in Equation~\eqref{eq:primal_suff} that replaces the balance constraint on the cluster-level sufficient statistics $\bar{S}_{G_i}$ with this average-to-one-within-cluster constraint.
Note that this average-to-one-within-cluster constraint will typically impose additional constraints: there are $K$ constraints --- one for each cluster --- rather than one constraint for each dimension of the cluster-level sufficient statistics $\bar{S}_{G_i}$ as in Equation~\eqref{eq:primal_suff}.
While these additional constraints may be redundant, constraining the weights to average to one within each cluster has the benefit of being robust to mis-specification of the cluster-specific term $\eta(s)$.
However, under this approach, we must again remove clusters that have only treated or only control units.
For the empirical analysis that follows, we implement both approaches, which we refer to as \textit{global balance (GB)} and \textit{average-to-one (AvTO)}.

\subsection{Summary of estimation methods}

We conclude this section with a brief summary. Above, we outlined that when we wish to adjust for clusters, we can target both global and local balance. Global balance captures the differences in treated and control units across all clusters, while local balance captures differences in treated and control units within each cluster. The most common approach, estimating the propensity score with a random intercept, does not include any constraint on local balance. Local balance can be targeted with random coefficients, but is generally infeasible due to computational issues. \citet{ben2024estimating} outline a strategy that regularizes over the local balance terms while ensuring that global balance holds. To our knowledge, their proposal is the only available method tailored to directly target both local and global balance. 

An alternative approach is to instead balance cluster level sufficient statistics. Here, we assume that clusters with similar covariates have a similar relationship with the outcome. This assumption allows us to overcome sparsity in the local balance constraints. We outlined a balancing weights framework for balancing these cluster-level sufficient statistics as well as their interactions with individual-level features.

To summarize the key trade-off: methods that condition on cluster membership (Section~\ref{sec:estimation_clust_membership}) require only Assumption~\ref{a:cluster_ignore}, which is nonparametric in cluster effects, but may struggle with small clusters or empirical overlap violations. The Mundlak approach (this section) can handle small clusters more gracefully but requires the stronger Assumption~\ref{a:exp_cluster_ignore}, which imposes additional structure on the data generating process. Researchers should choose between these approaches based on whether the exponential family assumption is plausible and whether cluster sizes permit direct conditioning on cluster membership.

Below, we conduct two empirical studies to review how to implement these methods and help readers understand the trade-offs between these different options.

\section{Statistical inference and bias correction}
\label{sec:inf}

We now turn to uncertainty quantification and statistical inference. We also introduce bias-correction, which augments the weighting estimator with an outcome model. Bias correction is useful when weighting fails to achieve good enough balance, but is also necessary to define our variance estimator. We again focus on estimating the ATT, noting that extensions to other estimands are straightforward.

Both variance estimation and bias correction depend on $\hat{m}_i$, which is an estimate of the expected outcome under control. Critically, there are three different covariate sets that vary with the identification assumptions. Specifically, under Assumption~\ref{a:ignore}, $\hat{m}_i  = \hat{m}(0, X_i)$, under Assumption~\ref{a:cluster_ignore}, $\hat{m}_i  = \hat{m}(0, X_i, G_i)$, and  under Assumption~\ref{a:exp_cluster_ignore}, $\hat{m}_i = \hat{m}(0, X_i, \bar{S}_{G_i})$. Under the AVTO approach, we additionally include cluster-specific fixed effects in the outcome model. Using the outcome model estimates $\hat{m}_i$, we can construct such a bias-corrected estimator as
\[
\hat{\mu}_0^\text{bc} = \frac{1}{n_1}\sum_{Z_i = 1} \hat{m}_i + \frac{1}{n_1}\sum_{Z_i = 0} \hat{\gamma}_i (Y_i - \hat{m}_i).
\]
In the appendix, we show that bias-correction has little effect on our empirical estimates and so we do not discuss it more here. Future work could more explicitly consider conditions under which such bias correction can be considered doubly robust.

Next, we review a variance estimator that also depends on $\hat{m}_i$. First, we write the squared standard error of the estimated treatment effect as $\hat{V}_\text{diff} = \hat{V}_1 + \hat{V}_0$, where $\hat{V}_1$ and $\hat{V}_0$ denote the estimated variance within the treated and control group, respectively. For the treated group, we can use the typical variance for the mean
$
\hat{V}_1 = \frac{1}{n_1^2}\sum_{Z_i = 1} \left(Y_i - \hat{\mu}_1\right)^2,
$
where $\hat{\mu}_1$ is a simple average of observed outcomes in the treated group. For the control group, this variance term is
\[
\hat{V}_0 = \frac{1}{\left(\sum_{Z_i = 0} \hat{\gamma}_i\right)^{2}}\sum_{Z_i = 0} \hat{\gamma}_i^2 \left(Y_i - \hat{m}_i\right)^2 = \sum_{Z_i = 0} \hat{\gamma}_i^2 \left(Y_i - \hat{m}_i\right)^2.
\]

\noindent We follow the terminology from \citet{ben2024estimating} and call $\hat{V}_0$ the residualized variance estimator (RVE). \citet{hirshberg2019minimax} and \citet{benmichael2021_review} provide technical conditions for the estimate $\hat{\mu}_0$ to be asymptotically normally distributed and for this variance estimate to be consistent. Crucially, those results require that the bias due to imbalance is asymptotically negligible relative to the standard error. Specifically, when weighting alone cannot achieve good enough balance, then the estimator requires additional bias correction via the outcome model.\footnote{Note that this variance estimator is for the in-sample ATT. To generalize to the population ATT we can add an additional term to $\hat{V}_0$, $\frac{1}{n_1^2}\sum_{Z_i = 1}(\hat{m}_i - \hat{\mu}_0)^2$, that measures the variability in the conditional expected potential outcome under control. However, this additional term will typically be small relative to $\hat{V}_1$ and so statistical conclusions about the sample will typically generalize to conclusions about the population.} With these variance estimates in hand, we can construct approximate $1-\alpha$ confidence intervals for the ATE via $\hat{\mu}_1 - \hat{\mu}_0 \pm z_{\alpha/2}\sqrt{\hat{V}_1 + \hat{V}_0}$, where $z_{\alpha/2}$ denotes the upper $\alpha/2$ quantile of a standard normal distribution.

A natural question is whether standard errors should be clustered at the group level. While the general advice in the literature is that standard errors should take into account clustering via a cluster robust variance estimator or the clustered bootstrap \citep{cafri2019review}, it is not necessary in this context. As \citet{abadie2023should} discuss, clustering corrections should be applied at the level of the data where treatment is assigned. In the current context, treatment is assigned to units \emph{within} clusters, and so there is no need to cluster the standard errors. That said, if the treatment were instead to have been assigned to clusters---schools or hospitals---instead of to students or patients within those clusters, then clustering corrections would be necessary \citep{ben2024approximate}.

\section{Simulation}
\label{sec:sim}

We now compare these approaches in a simulation study designed to answer two questions: (1) How does model-based IPW with a random intercept compare to methods that directly target local balance? (2) How do hierarchical balancing weights and Mundlak balancing weights differ in practice? 

We adapt the data generating process from \citet{salditt2023parametric}. We draw 10 individual-level covariates from a standard normal distribution; covariates $X_1, X_3, X_5, X_6, X_8$, and $X_9$ are then dichotomized to equal 1 when their value is $\geq 0$, and 0 otherwise. We also draw a cluster-level variable $U_g$ from a standard normal distribution. The propensity score $e_{ig}$ for individual $i$ in cluster $g$ is generated from the following model:
\begin{align}
e^\star_{ig} &= P(Z_{ig} = 1 \mid X_{1ig}, \ldots, X_{7ig}, U_g) = \frac{1}{1 + \exp[-f(X_{1ig}, \ldots, X_{7ig}, U_g)]} \\
e_{ig} &= 0.8 \cdot e^\star_{ig} + 0.15 
\end{align}
\noindent In the propensity score model, $f$ represents the following functional form:
\begin{align}
f(X_{1ig}, \ldots, X_{7ig}, U_g) &= 
\beta_0 + 0.8 X_{1ig} - 0.25 X_{2ig} + 0.6 X_{3ig} - 0.4 X_{4ig} - 0.8 X_{5ig} - 0.5 X_{6ig} + 0.7 X_{7ig} \nonumber \\
&\quad - 0.25 X_{2ig}^2 - 0.4 X_{4ig}^2 + 0.7 X_{7ig}^2 \nonumber \\
&\quad + 0.4 X_{1ig} X_{3ig} - 0.175 X_{2ig} X_{4ig} + 0.3 X_{3ig} X_{5ig} \nonumber \\
&\quad - 0.28 X_{4ig} X_{6ig} - 0.4 X_{5ig} X_{7ig} + 0.4 X_{1ig} X_{6ig} \nonumber \\
&\quad - 0.175 X_{2ig} X_{3ig} + 0.3 X_{3ig} X_{4ig} - 0.2 X_{4ig} X_{5ig} - 0.4 X_{5ig} X_{6ig} + \rho_U U_g \label{eq:ps-function}
\end{align}
Observed treatment assignments are then drawn from a Bernoulli distribution:
$
Z_{ig} \sim \text{Bernoulli}(e_{ig}).
$
\noindent Finally, the outcome model is
\begin{align}
Y_{ig} &= -3.85 + Z_{ig} + 0.3 X_{1ig} - 0.36 X_{2ig} - 0.73 X_{3ig} - 0.2 X_{4ig} \nonumber \\
&\quad + 0.71 X_{8ig} - 0.19 X_{9ig} + 0.26 X_{10ig} + \alpha_U U_g + \varepsilon_{ig} \label{eq:outcome}
\end{align}

\noindent where $\varepsilon_{ig} \sim \mathcal{N}(0,2)$ and the true treatment effect is $\tau = -0.4$. The covariates play different roles: $X_1$ to $X_4$ are confounders, $X_5$ to $X_7$ predict treatment only, and $X_8$ to $X_{10}$ predict the outcome only. Each simulated dataset contains 100 clusters with approximately 50 units per cluster. 

The key feature of this design is that $U_g$ acts as an unobserved cluster-level confounder. We vary the strength of this confounding by setting $\rho_U \in \{0, 0.25, 0.5\}$, with $\alpha_U = 0.5$ whenever $\rho_U > 0$. When $\rho_U = 0$, there is no cluster-level confounding and methods that account for clusters are unnecessary. When $\rho_U > 0$, cluster-based methods should outperform methods that ignore cluster membership.

We compare four methods:
\begin{itemize}
    \item \textbf{Balancing weights with no cluster adjustment.} A baseline that does not adjust for clusters.

    \item  \textbf{Model-based IPW with random intercept (RI-IPW).} The standard approach in the literature. As discussed in Section \ref{sec:model-ipw-global-local}, this targets global balance only. Incorporating random coefficients was infeasible in practice.

    \item \textbf{Hierarchical balancing weights.} The estimator from \citet{ben2024estimating}, described in Section \ref{sec:bal_wts_global_local}, which targets both global and local balance by conditioning on cluster membership.

    \item \textbf{Mundlak balancing weights.} The estimator from Section \ref{sec:mundlak_bal_wts}, which targets cluster-level sufficient statistics rather than cluster membership.
\end{itemize}
We implemented balancing weights using the \texttt{balancer} library in R \citep{benmichael2022} and measured the performance of each method via: (1) standardized absolute bias, $\left|\E[\widehat{\text{effect}} - \text{effect}]/\E[\text{effect}]\right|$; and (2) the root mean-squared error (RMSE), $\sqrt{\E[(\widehat{\text{effect}} - \text{effect})^2]}$.  We repeated the simulation 1,000 times.

\begin{figure}[tbp]
  \centering
    \includegraphics[scale=0.7]{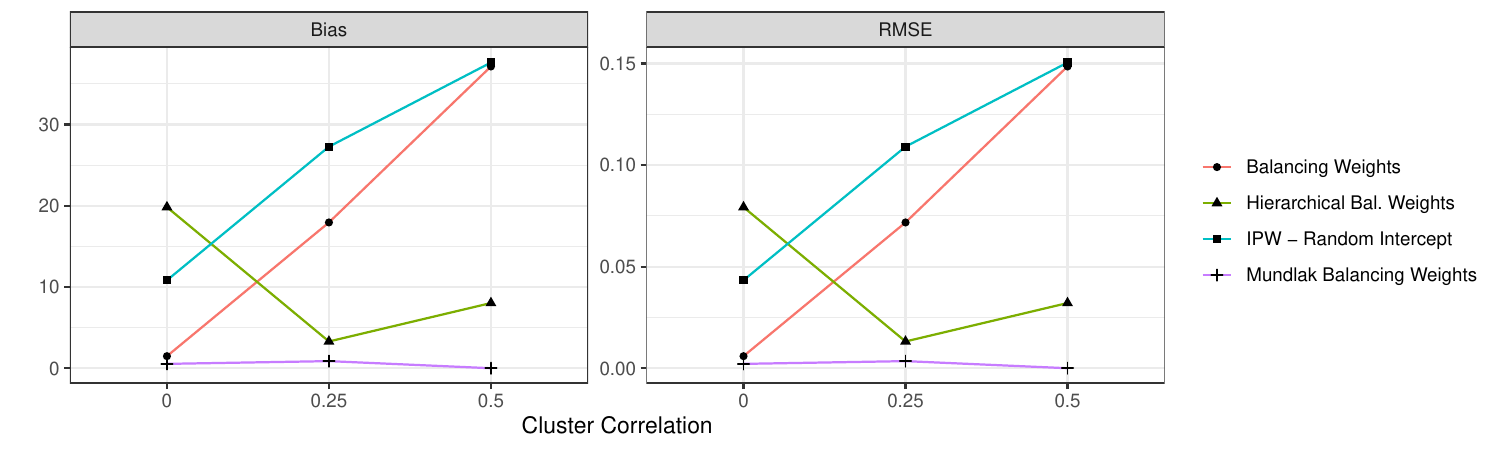}
     \caption{Bias and Root Mean Square Error of balancing weights, IPW based on a random intercept model, hierarchical balancing weights, and Mundlak weights, varying the degree of cluster confounding.}
  \label{fig:sims}
\end{figure}

Figure~\ref{fig:sims} presents the results. When there is no cluster-level confounding ($\rho_U = 0$), hierarchical balancing weights shows some bias and has the highest RMSE; RI-IPW is also biased and has the next highest RMSE. Thus, there is some cost to accounting for clusters when clusters are irrelevant. By contrast, standard balancing weights and Mundlak weights are both unbiased in this scenario---unsurprising for standard balancing weights, but notable for Mundlak weights.

Once we introduce cluster-level confounding, bias for standard balancing weights grows substantially, as expected. More surprising is that bias for RI-IPW also increases and actually exceeds the bias for standard balancing weights; this pattern persists at the highest level of confounding. At moderate confounding, hierarchical balancing weights and Mundlak weights perform nearly identically. At the highest level of confounding, bias for hierarchical balancing weights increases, while Mundlak weights remain nearly unbiased. 

These results yield three insights. First, RI-IPW performs poorly across all conditions. While some bias when clusters are irrelevant might be expected, the high bias when cluster-level confounding is present is striking: RI-IPW fails to outperform a method that ignores clusters entirely. Focusing on global balance alone does little to reduce bias from cluster-level confounding. Second, hierarchical balancing weights effectively account for cluster-level confounders but incur a cost when clusters are irrelevant. Third, Mundlak weights perform well across all scenarios: they remain nearly unbiased whether or not cluster-level confounding is present, avoiding the penalty that other cluster-based methods pay when clusters are irrelevant.

\section{Applications}
\label{sec:app}

We now present results from two applications with contrasting cluster structures: an education study examining pre-school programs and test scores, and a health services study examining emergency general surgery. In both cases, units are nested within clusters and the goal is to condition on clusters to capture unmeasured context. The applications differ in cluster size: the education application has many clusters but few units per cluster, while the health application has many clusters with widely varying sizes, including some quite large clusters.

For both applications, we compare four methods that adjust for observed confounders, as described in Section \ref{sec:sim}: (1) balancing weights with no cluster adjustment; (2) model-based IPW with a random intercept; (3) hierarchical balancing weights; and (4) Mundlak balancing weights. These comparisons allow us to assess the effect of cluster adjustment relative to individual-level balancing alone, and the added value of local balance constraints beyond RI-IPW, which only enforces global balance and is widely used in applied work. We also use agreement between hierarchical and Mundlak balancing weights as an informal check on the plausibility of the stronger Assumption~\ref{a:exp_cluster_ignore}: since Mundlak methods require this additional exponential family structure while hierarchical methods do not, substantial disagreement between these estimates could indicate that the chosen sufficient statistics do not fully capture cluster-level confounding.

We report select results in the main text. For the Mundlak methods, we implement two versions: \emph{Global balance (GB)}, which enforces global balance on cluster-level covariates, and \emph{Average-to-one (AvTO)}, which constrains weights to average to one within each cluster. Since we found limited differences between these two approaches, we report GB estimates in the main text and defer AvTO results to the supplementary materials. We also estimated treatment effects after bias-correction with outcome models, which made minor differences; these results also appear in the supplementary materials. Next, we review how we specify the basis functions. First, $\phi(x)$ is comprised of the unit level covariates, with splines fits added to those continuous covariates with nonlinear associations with the outcome. Next, for the Mundlak weights, the basis functions are more complex. Here, we condition on $\phi(x)$, $z\phi(x)$, and $\psi(s,x)$, where $\phi(x)$ is the set of unit level covariates, $z\phi(x)$ is set of the first order interactions between $\phi(x)$ and the proportion treated and control in each cluster, and $\psi(s,x)$ is the set of first order interactions between $\phi(x)$ and $\bar{S}_{G_i}$, which is vector the group means of the unit level covariates.

We use the \texttt{balancer} library in R \citep{benmichael2022} for all methods except model-based IP weights. We selected all hyperparameter values using a data driven method based on a regression of $Y$ on $X$ restricted to the control group $(Z=0)$. The hyperparameter is the variance of the residuals from this regression \citep{benmichael2022}. For each application, we estimate the ATT and associated 95\% confidence interval. We also report the effective sample size (ESS) for $\hat{\mu}_0$ (since we are targeting the ATT): methods that balance more features tend to have lower ESS and thus higher variance.

\paragraph{Balance metrics.}
We report several global and local balance metrics. First, we use a common metric known as the standardized mean difference for covariate $j$, $\text{SMD}_j$, and the cluster-specific version, $\text{SMD}_{g,j}$:
\[
\text{SMD}_{j} =  \frac{
            \bar\phi^{(1)}_{j} - \bar\phi^{(0)}_{j}
        }{
            s_j
        }
\qquad
\text{SMD}_{g,j} =  \frac{
            \bar\phi^{(1)}_{g,j} - \bar\phi^{(0)}_{g,j}
        }{
            s_{g,j}
        }
\]
\noindent where $\bar\phi^{(1)}_{j}$ is the mean in the treated group; $\bar\phi^{(0)}_{j}$ is the weighted mean in the control group; $s_j$ is the pooled standard deviation, $s_j = \sqrt{\frac{1}{2}\left[
            \left(s^{(1)}_{j}\right)^2 +
            \left(s^{(0)}_{j}\right)^2
        \right]}$; 
and $g$ denotes the cluster-specific versions.
We also compute an $L_2$ measure of imbalance across covariates both globally and locally:
\[
L_2^{\text{global}}
=
\sqrt{
    \frac{1}{p}
    \sum_{j=1}^{p} 
    \left(
       \text{SMD}_{j} 
    \right)^2
},\qquad\qquad
L_2^{\text{local}} = \sqrt{\frac{1}{PK}\sum_{j=1}^p\sum_{g=1}^K \text{SMD}_{g,j}^2}
\]

\noindent We can compare how each method reduces the amount of imbalance by calculating the percent reduction in the $L_2$ metric with and without weighting: we call this the percentage bias reduction (PBR). We should note that we calculate multiple $L_2$ values, since different methods balance different features of the covariates and as such are not always directly comparable.

\subsection{Education: Small Clusters}

The first application focuses on the impact of pre-school programs on post-kindergarten math test scores from \citet{dong2020using} and \citet{lee2021partially}. Here, students are nested in schools, and we wish to account for school context in the estimation of the treatment effects. That is, the analytic goal is to use schools to account for aspects of ``unmeasured context'' that affect test scores: school membership should capture many parts of the learning process that should be related to test scores.

The data are from the Early Childhood Longitudinal Study's Kindergarten (ECLS-K) study, which is a nationally representative cohort of children. In the data, there are 9,862 students, where 6,993 students attended a pre-school program and 2,869 did not.  Following \citet{lee2021partially}, we focus on the effect of \textit{not-attending} a pre-school program; that is, the ATT is the impact of not attending pre-school on students who did not attend pre-school.  There are 937 schools that serve as the cluster indicators. Student-level baseline covariates include sex, race, age at the start of kindergarten, height, weight, family type, family socio-economic status, income, parent's education level, and indicator for whether the student lives with their biological mother. The outcome variable of interest is the average math test score. Here, the cluster sizes are small and range from 1 to 22 students per school. There are 131 clusters that do not contain any treated students, and 27 schools that do not contain any control students. One key difference across the different estimation strategies we consider is their ability to include schools where all or none of the students are treated. As such, one relevant question is whether dropping these clusters---and changing the estimand---produces substantially different results. Removing schools with no treated or control students leaves 8,703 students nested in 779 schools.

\begin{table}[tbp]
\centering
\caption{Effective Sample Sizes for Five Different Estimators in the Class Size Application}
\label{tab.class.ecls}
\begin{tabular}{lc}
\toprule
Estimator & Effective Sample Size \\ 
\midrule
Unadjusted & 6,993 \\ 
Balancing weights with no cluster adjustment & 3,172 \\ 
Model-based IPW with cluster random intercepts & 4,662 \\ 
Hierarchical balancing weights & 1,133 \\ 
Mundlak balancing weights with GB constraint & 595 \\ 
\bottomrule
\end{tabular}
\end{table}

Table~\ref{tab.class.ecls} reports the effective sample size for $\hat\mu_0$ for each estimator. Before weighting, there are 6,993 control students. Standard balancing weights and IPW with a cluster random intercept have large effective sample sizes of 3,172 and 4,662, respectively. At the other extreme,  hierarchical and Mundlak balancing weights yield effective sample sizes that are much smaller: 1,133 and 595, respectively. In general, methods that only target global balance have a substantially higher ESS than those that also target local balance. This is not surprising, since targeting a much larger set of features typically requires more variable weights, which reduces the ESS.
However, note that the Mundlak approach balances fewer features than the hierarchical balancing approach (the dimension of $\psi(x,s)$ is 1200, versus the number of local balance features $d \times K = 18,696$). Despite this, the Mundlak approach has a lower ESS. This is primarily due to the scale and choice of the hyperparameter $\lambda$ being different between the two approaches.

\begin{figure}[tbp]
  \centering
    \includegraphics[scale=0.6]{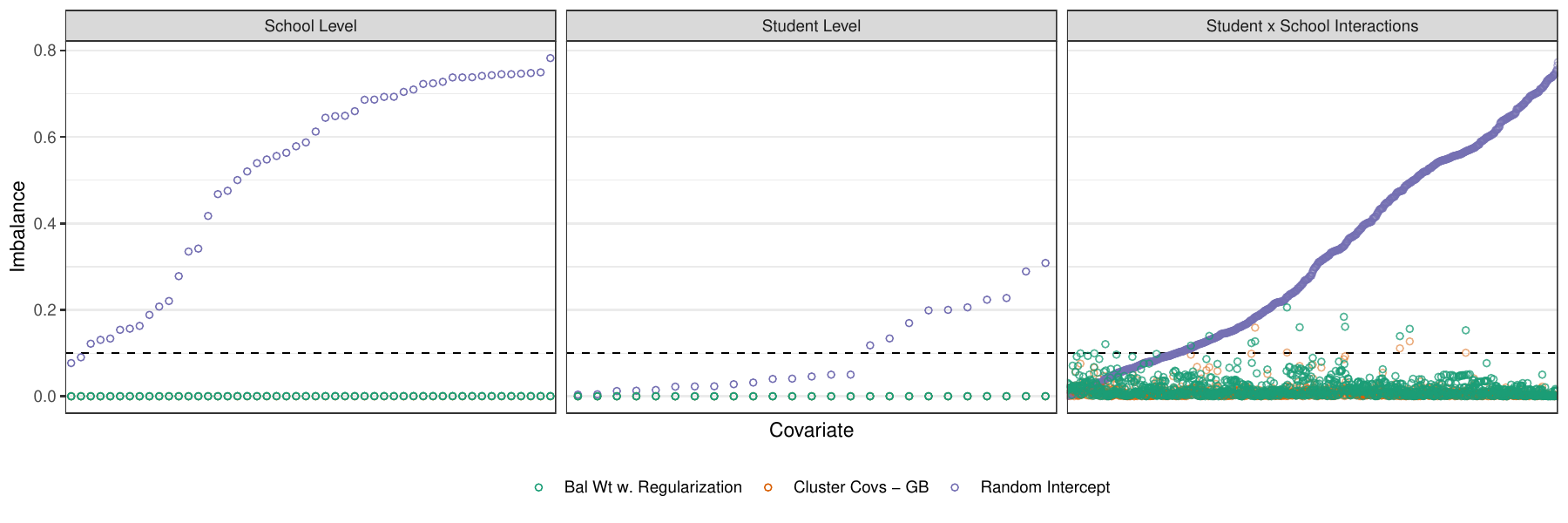}
     \caption{Balance metrics for three cluster adjustment methods in the education application.}
  \label{fig:ecls.bal.plot}
\end{figure}

Figure~\ref{fig:ecls.bal.plot} shows covariate imbalance for the three cluster adjustment methods for covariates defined at the school level, student level, and student $\times$ school interactions. As expected, random intercept IPW fails to balance school-level covariates and also allows for substantial local imbalance. The two balancing weight methods designed for clusters largely reduce these imbalances while also improving school-level covariate balance. The aggregated $L_2$ imbalance measures reflect these qualitative comparisons. In the full data set (before removing any clusters), the unadjusted global imbalance metric is $L_2^{\text{global}} = 0.33$; RI-IPW reduces this to $L_2^{\text{global}} = 0.14$; both standard and hierarchical balancing weights reduce this even further to $L_2^{\text{global}} = 0.001$. Thus, balancing weight approaches produce far better global balance than RI-IPW. The corresponding unadjusted local imbalance metric is $L_2^{\text{local}} = .26$; hierarchical balancing weights reduces this to $L_2^{\text{local}} = .05$, a reduction of 80\%, while RI-IPW does not meaningfully change this metric at all.  Thus, consistent with the results in Figure~\ref{fig:surg.bal.plot}, we see dramatic gains in covariate balance for balancing weights methods relative to RI-IPW. 

\begin{figure}[tbp]
  \centering
    \includegraphics[scale=0.6]{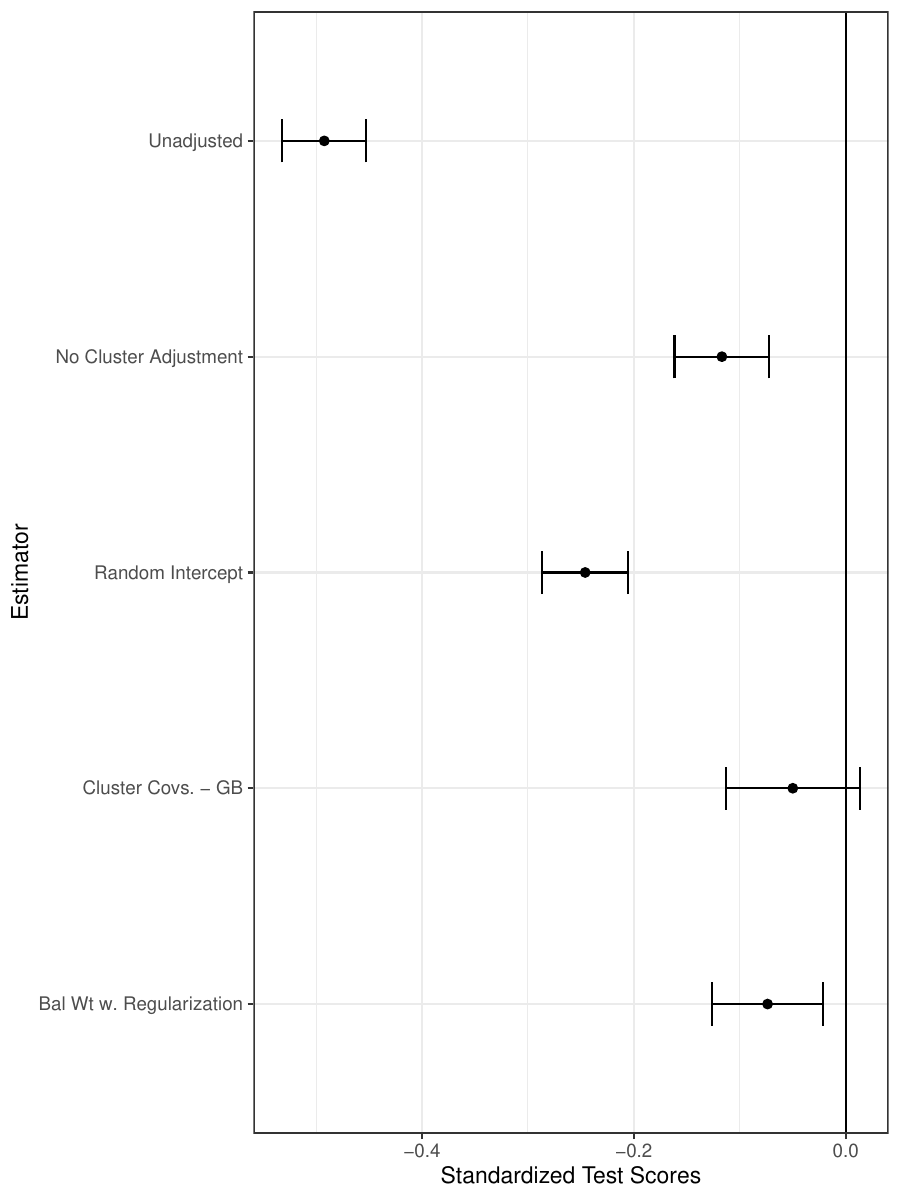}
     \caption{Estimates for the effect of not attending a pre-school program on test scores for four different covariate adjustment strategies.}
  \label{fig:ecls.plot}
\end{figure}

Figure~\ref{fig:ecls.plot} summarizes results from the four estimation methods. Without any adjustment, students who did not attend a pre-school program had 0.5 SD lower math scores in kindergarten compared to their peers who attended preschool, a substantial difference. Adjusting for student-level covariates alone reduces this difference to 0.12 SD, with confidence intervals that exclude zero. Adjusting for cluster membership through RI-IPW returns a treatment effect of 0.25 SD. Here, the treatment effect seems rather large, likely reflecting the substantial imbalance left by RI-IPW. This estimate is nearly identical to those in \citet{lee2021partially}, which takes a different a different approach to cluster covariates, but also does not include any local balance constraints. Hierarchical balancing weights yields a point estimates around 0.07 SD, with confidence intervals that also exclude zero. Finally, the Mundlak balancing weight estimate is also 0.07 SD, but the confidence intervals do include zero, consistent with the much lower ESS reported in Table \ref{tab.class.ecls}. Notably, discarding smaller clusters appears to make little difference across these approaches. 

Thus, there are meaningful practical differences in cluster adjustment methods. RI-IPW produces much larger  in magnitude than hierarchical balancing weights or Mundlak balancing weights. Given the similarity between the balancing weights estimates and the substantial imbalances in Figure \ref{fig:ecls.bal.plot}, this larger estimate likely reflects a failure to balance key covariates.
At the same time, both hierarchical and Mundlak balancing weights produce nearly identical estimates that are much smaller than the other estimates. This stresses the importance of incorporating a local balance constraint or more fully conditioning on cluster covariates, albeit at the cost of smaller effective sample sizes. The agreement between hierarchical and Mundlak estimates is reassuring: recall that hierarchical balancing weights require only Assumption~\ref{a:cluster_ignore}, while Mundlak weights require the stronger Assumption~\ref{a:exp_cluster_ignore}. The similarity of these estimates suggests that the exponential family assumption underlying the Mundlak approach is not grossly violated in this application---although it does not guarantee that the assumption holds exactly.


\subsection{Health Services Research: Large Clusters}

Emergency general surgery (EGS) refers to medical emergencies where the injury is internal (e.g., a burst appendix); there are currently 51 medical conditions with this designation. EGS conditions account for more than 800,000 operations with an estimated 3-4 million patients annually in the United States \citep{shafi2013emergency,gale2014public,havens2015excess,ogola2015financial,scott2016use}. Much research has focused on the comparative effectiveness of operative versus non-operative care for EGS conditions, and has found that the effectiveness of operative care varies with EGS conditions \citep{hutchings2022effectiveness,kaufman2023operative,moler2022local,grieve2023clinical}. Here, we focus on whether operative treatment has benefits for patients with hepato-pancreato-biliary (HPB) disorders, which are conditions that affect the liver, pancreas, and biliary system. Common HPB disorders include gallstones, cholecystitis (an inflammation of the gallbladder), and pancreatitis (an inflammation of the pancreas).

We use a dataset that merges the American Medical Association (AMA) Physician Masterfile with all-payer hospital discharge claims from New York, Florida, and Pennsylvania in 2012-2013. The study population includes all patients admitted for emergency or urgent inpatient care for HPB disorders. The data includes a small set of demographic measures, such as indicators for race and ethnicity, sex, age, and insurance type. The dataset also includes measures of baseline patient frailty, represented by an indicator of severe sepsis or septic shock and pre-existing disability. There are indicators for 31 comorbidities based on Elixhauser indices \citep{elixhauser1998comorbidity}, which represent prior medical conditions that may complicate treatment for HPB conditions. Finally, there are three indicators for the three types of surgery used to treat HPB cases. We include these indicators variables as well to control for variation in surgery type. The primary outcome is a binary measure for the presence of an adverse event following either treatment, i.e., death or a prolonged length of stay in the hospital.

In this setting, hospitals are the relevant cluster, since we are concerned that there are key unmeasured aspects of care that differ from hospital to hospital. In other words, we imagine that the ideal comparison is between patients that received surgery and those who did not within the sample hospital, which would hold constant hospital-level systemic factors. In our data, there are 143,415 patients nested within 540 hospitals. Patient population sizes within hospital vary considerably from a single HPB patient to 1,556 patients per hospital, with a mean of 265 patients. There are 10 hospitals with no treated patients and 18 hospitals in which every patient is treated. Removing these 28 hospitals removes 176 patients, or 0.1 percent of all patients.

Table~\ref{tab.surg.ess} contains the effective sample size for estimating $\hat\mu_0$. There are 36,786 control patients before weighting. Balancing weights with no cluster adjustment reduces the ESS by half (19,348); incorporating a random hospital intercept restores most of the ESS (31,124). Thus, targeting global balance alone has only a modest impact on the effective sample size. 
By contrast, additionally targeting local balance dramatically reduces ESS, as in the education example above: hierarchical and Mundlak weights have effective sample sizes of 4,775 and 6,623, respectively.

\begin{table}[tbp]
\centering
\caption{Effective Sample Sizes for Five Different Estimators in the Surgery Application}
\label{tab.surg.ess}
\begin{tabular}{lc}
\toprule
Estimator & Effective Sample Size \\ 
\midrule
Unadjusted & 36,786 \\ 
Balancing weights with no cluster adjustment & 19,348 \\ 
Model-based IPW with cluster random intercepts & 31,124 \\ 
Hierarchical balancing weights & 4,775 \\ 
Mundlak balancing weights with GB constraint & 6,623 \\ 
\bottomrule
\end{tabular}
\end{table}

\begin{figure}[htbp]
  \centering
    \includegraphics[scale=0.6]{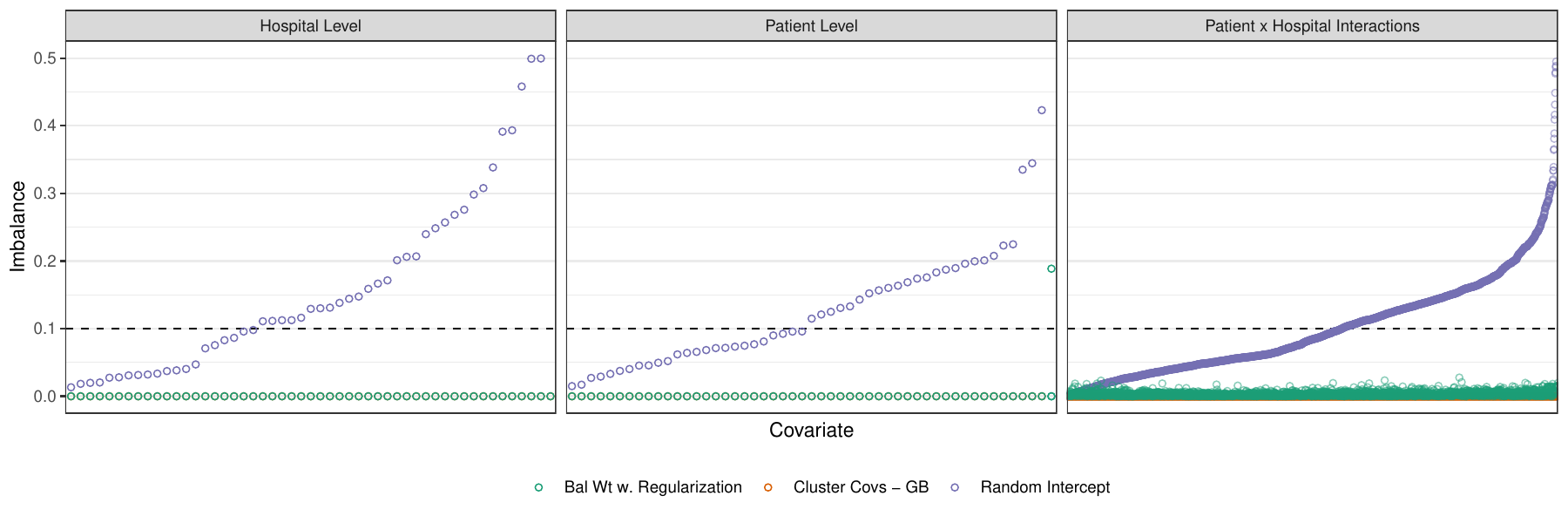}
     \caption{Balance metrics for three cluster adjustment methods in the surgery application.}
  \label{fig:surg.bal.plot}
\end{figure}

Figure~\ref{fig:surg.bal.plot} shows covariate imbalance for the three cluster adjustment methods for covariates defined at the hospital level, patient level, and patient $\times$ hospital interactions. Much like in the education application, RI-IPW fails to balance hospital-level covariates and also allows for substantial local imbalance. The two balancing weight methods designed for clusters completely reduce these imbalances while also improving hospital-level balance.

As in the education example, the aggregated $L_2$ imbalance measures reflect these qualitative comparisons. In the full data set (before removing any hospitals), the unadjusted global imbalance metric is $L_2^{\text{global}} = 0.22$; RI-IPW reduces this to $L_2^{\text{global}} = 0.15$; both standard and hierarchical balancing weights again reduce this to $L_2^{\text{global}} = 0.001$. Thus, balancing weights approaches produce far better global balance than RI-IPW. The corresponding unadjusted local imbalance metric is $L_2^{\text{local}} = .21$; hierarchical balancing weights reduces this to $L_2^{\text{local}} = .02$, a reduction of 89\%, while RI-IPW does not meaningfully change this metric at all. Again, we see dramatic gains in covariate balance for balancing weights methods relative to RI-IPW.

\begin{figure}[htbp]
  \centering
    \includegraphics[scale=0.6]{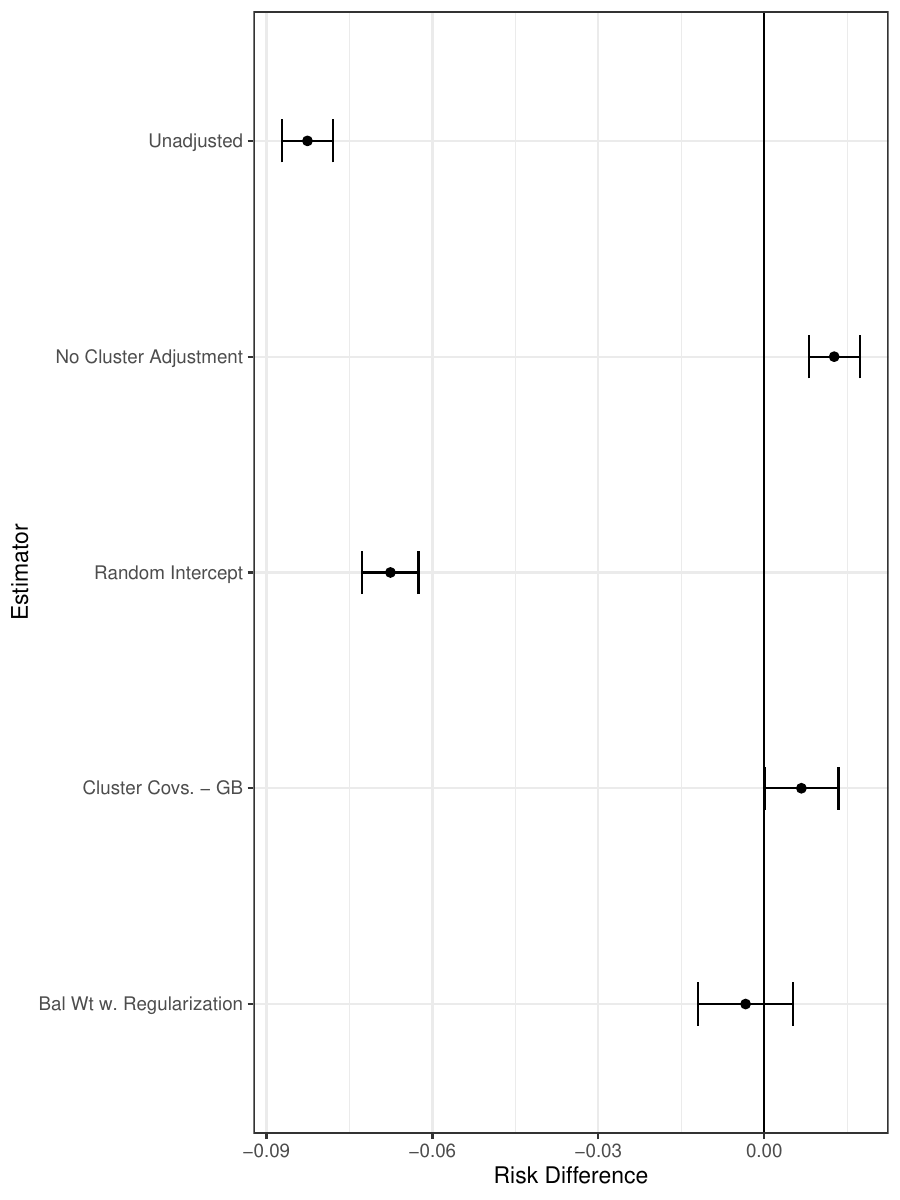}
     \caption{Estimates for the effect of surgery on adverse events for five different covariate adjustment strategies.}
  \label{fig:surg.plot}
\end{figure}

Figure~\ref{fig:surg.plot} shows the point estimates and 95\% confidence intervals for the effect of surgery.  The unadjusted estimate indicates that surgical care lowers the risk of an adverse event by almost 9\%, although this is likely biased if healthier patients are selected for surgery. Indeed adjusting for patient covariates alone flips the sign and indicates that surgical care actually increases the risk of an adverse event by around 3\%, with confidence intervals bounded away from zero. 
Next, RI-IPW yields an estimate that is quite close to the unadjusted estimate. However, as we observed above, RI-IPW fails to achieve good covariate balance and this estimate likely reflects substantial bias. Hierarchical balancing weights, by contrast, estimate an effect that is essentially zero---and is much closer to the balancing weights estimate that does not include cluster adjustments. Finally, adjusting for hospital-level sufficient statistics in the Mundlak estimator similarly yields a small estimate that is not statistically significant.

Comparing these estimates, RI-IPW likely yields a poor estimate of the causal effect, with large covariate imbalance and with estimates that are clear outliers relatively to the other methods; this is consistent with both the simulation evidence and the education application. Next, the two cluster adjustments methods produce similar estimates, which indicates that accounting for the hospital context is likely an important part of the design. We find that there appears to be little benefit to surgical treatment for HPB conditions, once we account for the care environment. By contrast, if we only balance patient covariates, we would find that surgery appears to harm patients.

As in the education application, the agreement between hierarchical and Mundlak balancing weights provides some reassurance about Assumption~\ref{a:exp_cluster_ignore}. However, researchers should note that this agreement does not \emph{validate} the stronger assumption; it merely suggests that the sufficient statistics capture enough cluster-level variation to produce similar estimates. In settings where hierarchical methods are feasible---as they are here given the larger cluster sizes---researchers may prefer hierarchical balancing weights precisely because they avoid the additional exponential family assumption.

\section{Discussion}

Taken together, our results suggest that researchers should generally prefer balancing weights approaches to traditional, model-based IPW with a hierarchical propensity score. RI-IPW is particularly problematic when treatment assignment varies within clusters. When feasible, researchers should also prioritize methods that target both global and local balance, such as hierarchical and Mundlak balancing weights. 
The latter are especially promising when cluster-level covariates are available and when cluster sizes are small. However, researchers should recognize that Mundlak methods require the stronger Assumption~\ref{a:exp_cluster_ignore}, which posits that cluster-level sufficient statistics fully capture cluster-level confounding. When cluster sizes are large enough to permit direct conditioning on cluster membership, hierarchical balancing weights under the weaker Assumption~\ref{a:cluster_ignore} may be preferable, as they avoid the additional exponential family structure.

In general, we recommend comparing estimates across multiple methods (e.g., no cluster adjustment, hierarchical balancing weights, and Mundlak balancing weights). If cluster adjustment is important, estimates from methods that target local balance should be similar to each other and potentially different from estimates that ignore clusters or only target global balance. Agreement between hierarchical and Mundlak balancing weights provides some reassurance that Assumption~\ref{a:exp_cluster_ignore} is not grossly violated; disagreement may indicate that the sufficient statistics do not fully capture cluster-level confounding or that the hierarchical approach was unable to achieve good local balance.

There are several important directions for future research. 
First, our paper focuses on weighting methods and only briefly addresses bias correction with an outcome model. Understanding the interplay between weighting and outcome modeling is important for this setting. 
Moreover, our focus has been on observational studies with a clustered data structure. \citet{arkhangelsky2024fixed}, however, connect this same setup to the panel data setting; similarly adapting our framework to panel data estimation is a promising direction. Another extension is to move beyond within-study average treatment effects, such as the ATT, to heterogeneous treatment effects and transportability.  Finally, we focus on only two levels of hierarchy here. While this is by far the most common setting, cases with multiple levels, such as students nested within classrooms nested within schools, introduce many additional complications.

\clearpage

\appendix

\singlespacing

\section{Additional Empirical Results}

Here, we present additional empirical results omitted from the main paper. We include two additional sets of results. First, we report the results from the AvTO Mundlak weighting approach. Second, we report treatment effect estimates that include augmentation from an outcome model. In both cases, the results are unchanged from those reported in the main text.

\subsection{Education: Small Clusters}

\begin{table}[htbp]
\centering
\caption{Effective Sample Sizes for Five Different Estimators in the Education Application}
\label{tab.surg.ess}
\begin{tabular}{lc}
\toprule
Estimator & Effective Sample Size \\ 
\midrule
Unadjusted & 6,993 \\ 
Balancing weights with no cluster adjustment & 3,173 \\ 
Model-based IPW with cluster random intercepts & 4,662 \\ 
Hierarchical balancing weights & 1,133 \\ 
Mundlak balancing weights with GB constraint & 595 \\ 
Mundlak balancing weights with AvTO constraint & 609 \\ 
\bottomrule
\end{tabular}
\end{table}

\begin{figure}[htbp]
  \centering
    \includegraphics[scale=0.6]{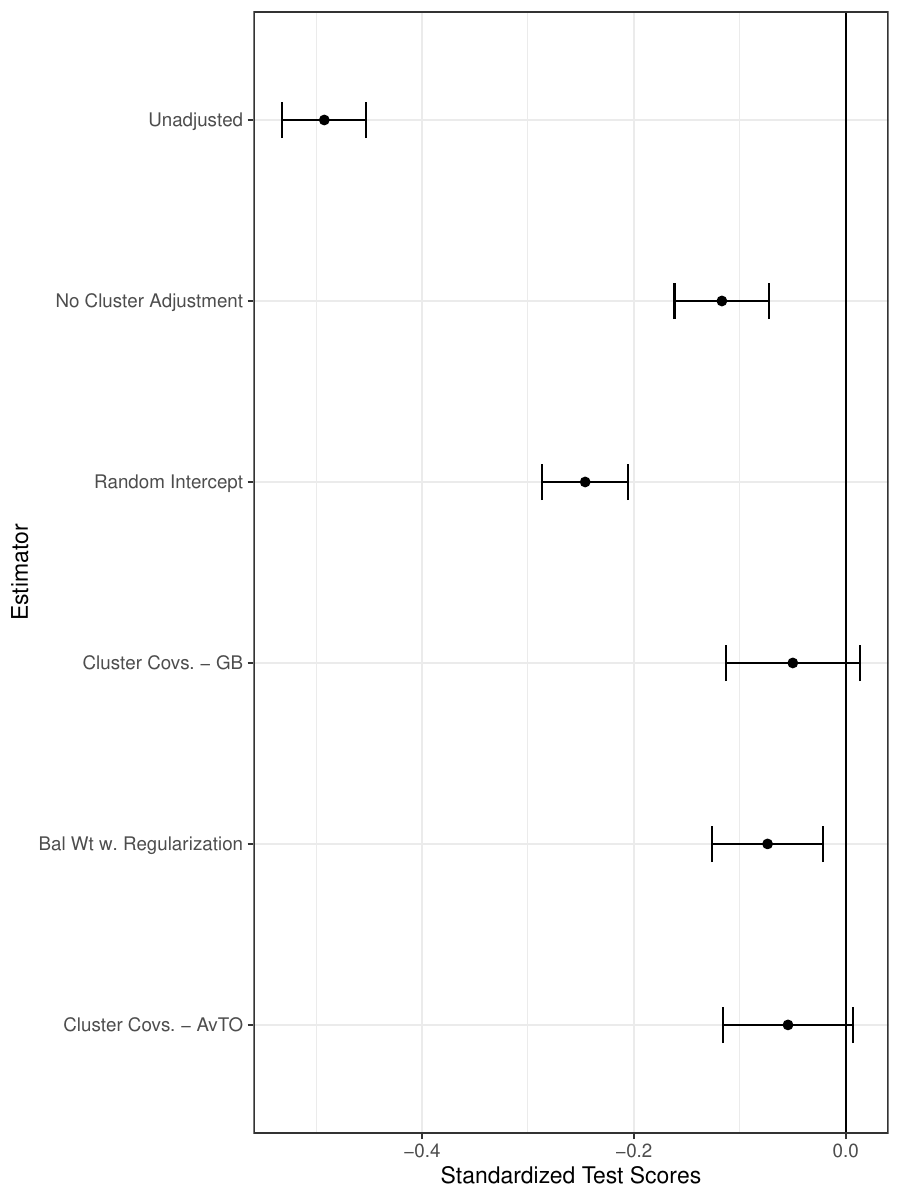}
     \caption{Estimates for the effect of not attending a pre-school program on test scores for five different covariate adjustment strategies.}
  \label{fig:ecls.plot.five}
\end{figure}

\begin{figure}[htbp]
  \centering
    \includegraphics[scale=0.6]{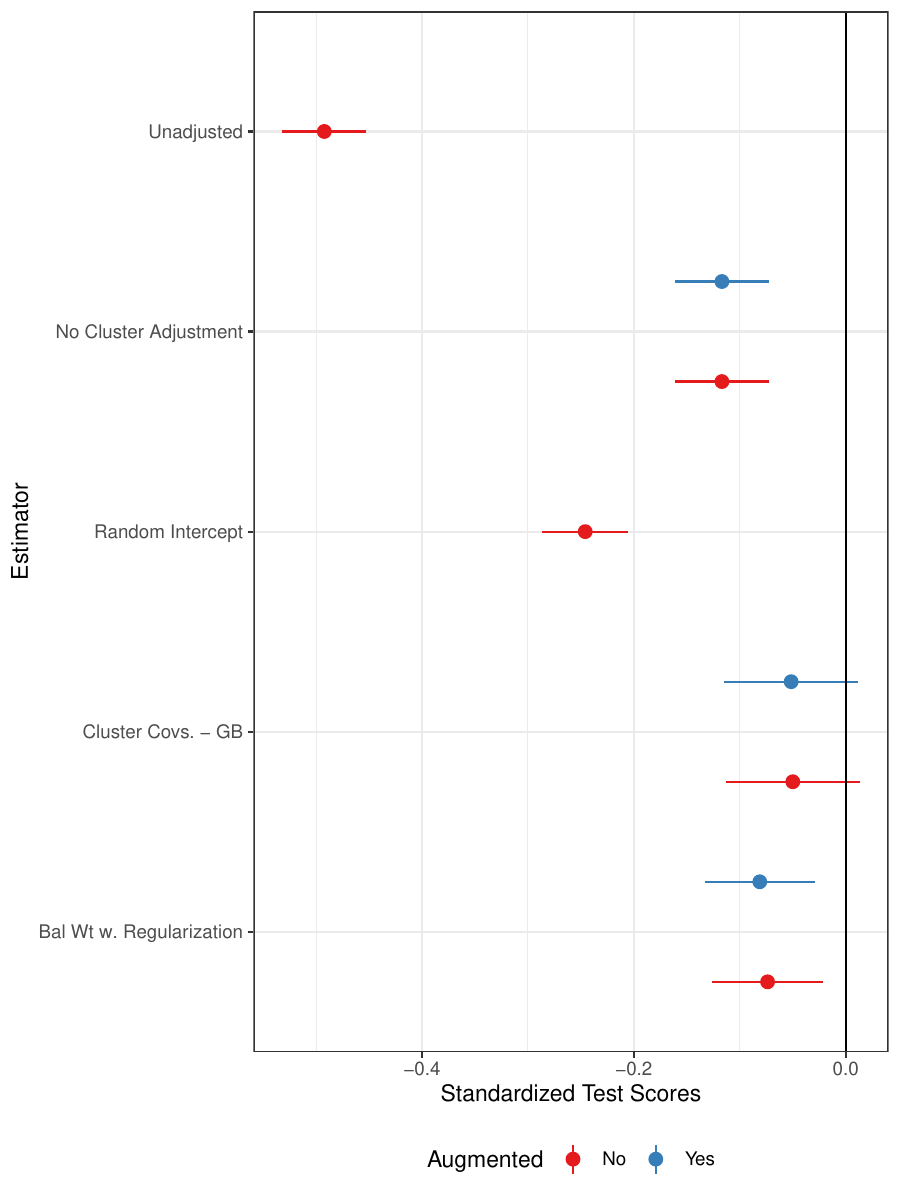}
     \caption{Estimates for the effect of not attending a pre-school program on test scores for five different covariate adjustment strategies with and without outcome model augmentation/bias correction.}
  \label{fig:ecls.plot.aug}
\end{figure}

\clearpage

\subsection{Surgery}

\begin{table}[htbp]
\centering
\caption{Effective Sample Sizes for Five Different Estimators in the Surgery Application}
\label{tab.surg.ess}
\begin{tabular}{lc}
\toprule
Estimator & Effective Sample Size \\ 
\midrule
Unadjusted & 36,786 \\ 
Balancing weights with no cluster adjustment & 19,348 \\ 
Model-based IPW with cluster random intercepts & 31,124 \\ 
Hierarchical balancing weights & 4,775 \\ 
Mundlak balancing weights with GB constraint & 6,623 \\ 
Mundlak balancing weights with AvTO constraint & 5,677 \\ 
\bottomrule
\end{tabular}
\end{table}

\begin{figure}[htbp]
  \centering
    \includegraphics[scale=0.6]{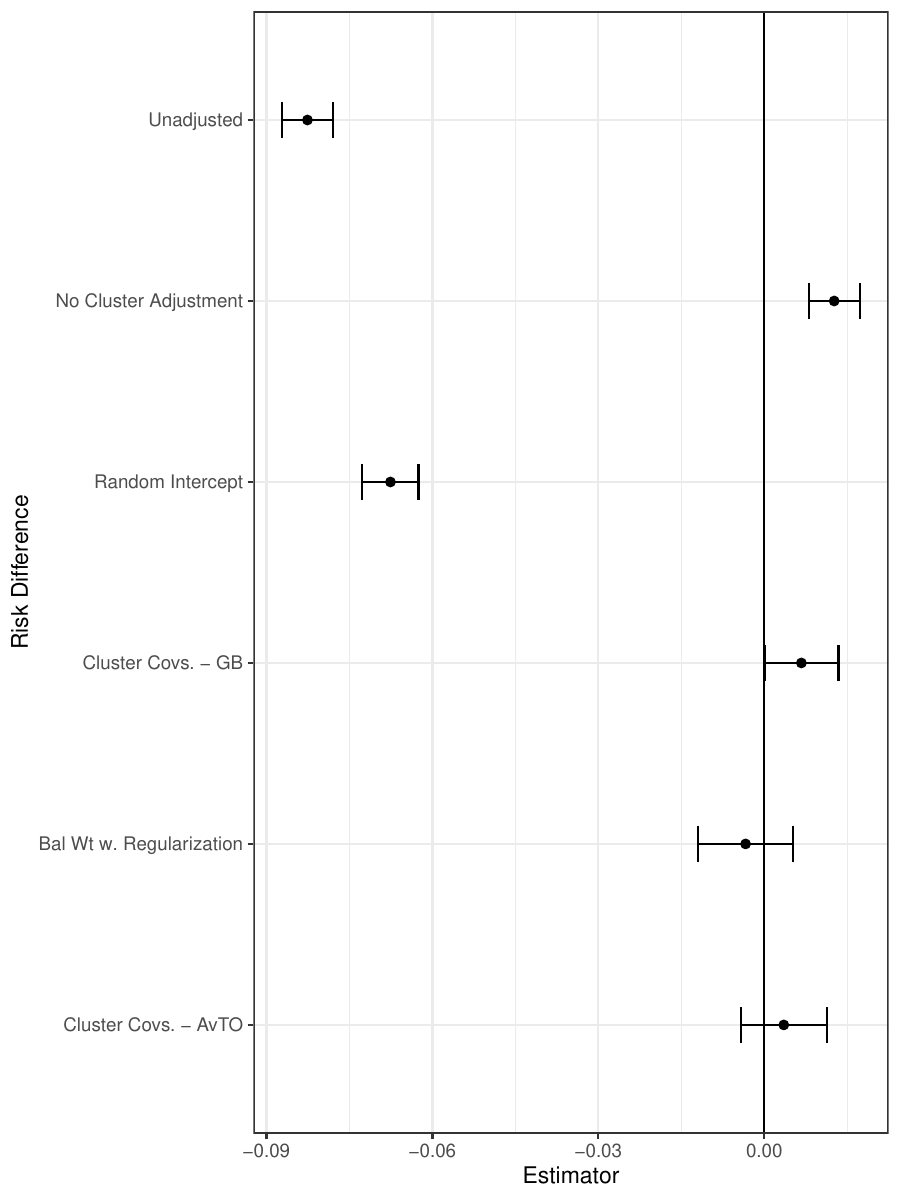}
     \caption{Estimates for the effect of surgery on adverse events for five different covariate adjustment strategies.}
  \label{fig:surg.plot.five}
\end{figure}

\begin{figure}[htbp]
  \centering
    \includegraphics[scale=0.6]{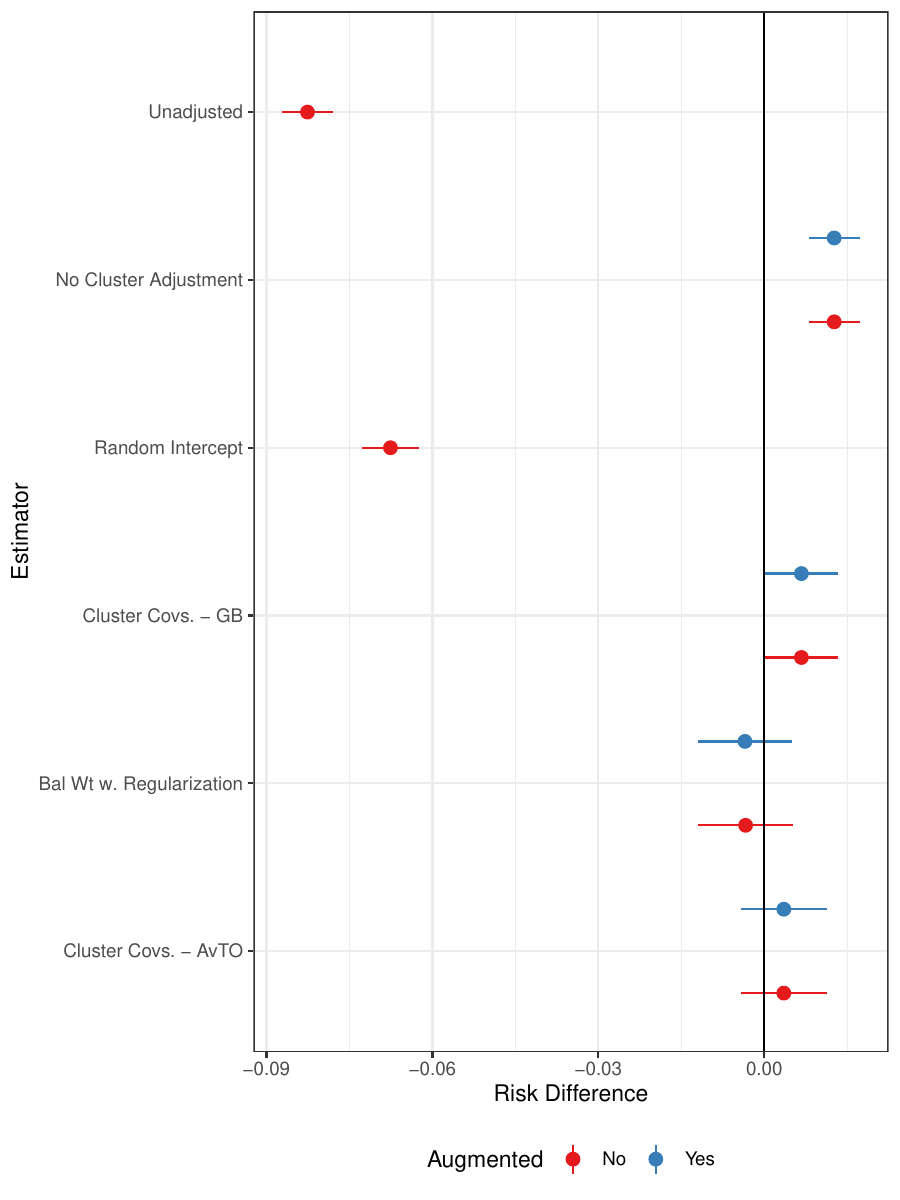}
     \caption{Estimates for the effect of surgery on adverse events for five different covariate adjustment strategies with and without outcome model augmentation/bias correction.}
  \label{fig:surg.plot.aug}
\end{figure}

\clearpage

\singlespacing
\bibliographystyle{asa}
\bibliography{bal-weight}

\end{document}